\documentclass[useAMS,usenatbib,aas_macros]{mnras}
\usepackage{graphicx}
\usepackage{amssymb}
\usepackage{footnote}

\usepackage{epsfig}
\usepackage{color}
\usepackage{amssymb}
\usepackage{ams math}
\usepackage[greek,english]{babel}
\usepackage{subfigure}
\usepackage{relsize}
\usepackage{xfrac}
\usepackage{placeins,lscape}
\usepackage{makecell}

\def \asec{$^{\prime\prime}$}

\def\civ{\mbox{C\,{\sc iv\sc{$\lambda$1548}}}}
\def\siiv{\mbox{Si\,{\sc iv\sc{$\lambda$1400}}}}
\def\nv{\mbox{N\,{\sc v\sc{$\lambda$1240}}}}
\def\niv{\mbox{N\,{\sc iv\sc{$\lambda$1720}}}}

\def\hii{\mbox{H\,{\sc ii}}}
\def\hi{\mbox{H\,{\sc i}}}

\def\hbeta{\mbox{H\,{\sc $\beta$}}}

\def\lya{\mbox{Ly\,{\sc $\alpha$}}}

\def\feiii{\mbox{Fe\,{\sc iii}}}
\def\fev{\mbox{Fe\,{\sc v\sc{$\lambda$1363}}}}
\def\sIiii{\mbox{Si\,{\sc iii\sc{$\lambda$1417}}}}
\def\ciii{\mbox{C\,{\sc iii\sc{$\lambda\lambda$1426/28}}}}
\def\sv{\mbox{S\,{\sc v\sc{$\lambda$1502}}}}

\def\oiii{\mbox{[O\,{\sc iii}]}}
\def\oiiiaur{\mbox{[O\,{\sc iii]\sc{$\lambda$4363}}}}
\def\oiiiauruv{\mbox{O\,{\sc iii]\sc{$\lambda$1661,1666}}}}

\def\oii{\mbox{[O\,{\sc ii}]}}

\def\fiby{\textsc{FiBY}}
\def\simba{\textsc{Simba}}

\def\logm{log($M_{\ast}$/M$_{\odot}$)}
\def\logz{log($Z_{\ast}$/Z$_{\odot}$)}

\defcitealias{weinberg2017}{WAF}
\defcitealias{leitherer2011}{L11}
\defcitealias{zahid2017}{Z17}
\defcitealias{steidel2016}{S16}

\newcommand{\Gyr}{\,{\rm Gyr}}
\newcommand{\Myr}{\,{\rm Myr}}

\title[Stellar metallicities at $z=2.5-5$]{The VANDELS survey: the stellar metallicities of star-forming galaxies at $\mathbf{2.5 < z < 5.0}$}
\author[F. Cullen et al.]{F. Cullen$^{1}$\thanks{E-mail:fc@roe.ac.uk}, 
R. J. McLure${^{1}}$, J. S. Dunlop${^{1}}$, S. Khochfar${^{1}}$, R. Dav\'e${^{1,2,3}}$, R. Amor\'in${^{4,5}}$, \and M. Bolzonella${^{6}}$, A. C. Carnall${^{1}}$, M. Castellano${^{7}}$, A. Cimatti${^{8,9}}$, M. Cirasuolo${^{10}}$, \and G. Cresci${^{9}}$, J. P. U. Fynbo${^{11}}$, F. Fontanot${^{12}}$, A. Gargiulo${^{13}}$, B. Garilli${^{13}}$, L. Guaita${^{7,14}}$, \and N. Hathi${^{15}}$, P. Hibon${^{16}}$, F. Mannucci${^{9}}$, F. Marchi${^{7}}$, D. J. McLeod${^{1}}$, L. Pentericci${^{7}}$, \and L. Pozzetti${^{6}}$, A. E. Shapley${^{17}}$,  M. Talia${^{6,8}}$ and G. Zamorani${^{6}}$\\
\\
Affiliations are listed at the end of the paper}

\begin{document}

\date{Accepted -- . Received 2017 May 30}

\pagerange{\pageref{firstpage}--\pageref{lastpage}} \pubyear{2016}

\maketitle	

\label{firstpage}

\begin{abstract}
We present the results of a study utilising ultra-deep, rest-frame UV, spectroscopy to quantify the relationship between stellar mass and stellar metallicity for 681 star-forming galaxies at $2.5<z<5.0$ ($\langle z \rangle = 3.5 \pm 0.6$) drawn from the VANDELS survey.
Via a comparison with high-resolution stellar population synthesis models, we determine stellar metallicities ($Z_{\ast}$, here a proxy for the iron abundance) for a set of high signal-to-noise ratio composite spectra formed from subsamples selected by mass and redshift.
Across the stellar mass range $8.5 < \mathrm{log}(\langle M_{\ast} \rangle/\rm{M}_{\odot}) < 10.2$ we find a strong correlation between stellar metallicity ($Z_{\ast}/\mathrm{Z}_{\odot}$) and stellar mass, with stellar metallicity monotonically increasing from $Z_{\ast}/\mathrm{Z}_{\odot} < 0.09$ at $\langle M_{\ast} \rangle = 3.2 \times 10^{8} \rm{M}_{\odot}$ to  $Z_{\ast}/Z_{\odot} = 0.27$ at $\langle M_{\ast} \rangle = 1.7 \times 10^{10} \rm{M}_{\odot}$.
In contrast, at a given stellar mass, we find no evidence for significant 
metallicity evolution across the redshift range of our sample.
However, comparing our results to the $z=0$ stellar mass-metallicity relation for star-forming galaxies, we find that the $\langle z \rangle = 3.5$ relation is consistent with being shifted to lower metallicities by $\simeq 0.6$ dex at all stellar masses.
Contrasting our derived stellar metallicities with estimates of the gas-phase metallicities of galaxies at similar redshifts and stellar masses, we find evidence for enhanced $\rm{O}/\rm{Fe}$ ratios in $z \gtrsim 2.5$ star-forming galaxies of the order (O/Fe) $\gtrsim 1.8$ $\times$ (O/Fe)$_{\odot}$.
Finally, by comparing our results to the predictions of three cosmological simulations, we find that the $\langle z \rangle = 3.5$ stellar mass-metallicity relation is consistent with current predictions for how outflow strength scales with galaxy stellar mass.
This conclusion is supported by an analysis of one-zone analytic chemical evolution models, and suggests that the mass loading parameter ($\eta=\dot{M}_{\mathrm{outflow}}/M_{\ast}$) scales as $\eta \propto M_{\ast}^{\beta}$ with $\beta \simeq -0.4$.
\end{abstract} 



\section{Introduction}

The relationship between the stellar mass and metallicity of galaxies as a function of cosmic epoch provides a fundamental constraint on models of galaxy formation \citep[e.g.][]{dere_metallica}.
The metallicity of a galaxy is determined by the integrated past history of star formation, the fraction of metal-enriched gas lost from the interstellar medium (ISM) through outflows, and the dilution of enriched ISM gas by pristine inflows from the intergalactic medium (IGM).
Metallicity scaling relations therefore provide rich information regarding some of the key physical processes governing the evolution of galaxies.

In recent years, significant efforts have been made towards determining the gas-phase metal content of galaxies at $2 \lesssim z \lesssim 4$ from the ratios of strong optical nebular emission lines \citep[e.g.][]{cullen2014,maier2014,steidel2014,troncoso2014,wuyts2014,salim2015,sanders2015a,onodera2016,kashino2017,sanders2018}.
Despite this, the substantial uncertainties inherent in converting nebular line ratios to abundances have frustrated efforts to reach a consensus.
In addition to the known inconsistency between different metallicity calibrations in the local Universe \citep[e.g.][]{kewley2008, barrera2017,sanchez2019}, the problem is compounded at higher redshifts by the evolution in the physical conditions in \hii \ regions, potentially rendering local calibrations unusable \citep{cullen2016,shapley2014,steidel2014,strom2017}.
Disconcertingly, it is still unclear whether the correlation between nebular line ratios and galaxy stellar mass in high-redshift galaxies is dictated primarily by the gas-phase metallicity or is purely driven by the ionizing properties of the massive stellar population \citep[e.g.][]{steidel2014}.
To definitively address these concerns, direct estimates of gas-phase metallicity are needed at high-redshift, requiring the the detection of faint oxygen auroral lines such as \oiiiaur \ and \oiiiauruv, which are extremely challenging observations that have only been achieved a handful of times at $z\gtrsim2$ \citep[e.g.][]{sanders2016,amorin2017}.

An alternative method for probing the metal content of galaxies utilizes their stellar continuum emission directly.
Abundances derived in this way are referred to as stellar $(Z_{\ast})$ as opposed to gas-phase $(Z_{\rm{g}})$ metallicities.
When available, stellar metallicities are a useful independent probe of the metal content of galaxies. 
In the local Universe, several authors have investigated the stellar mass-metallicity relationship for large statistical galaxy samples, primarily from the extensive Sloan Digital Sky Survey (SDSS) dataset \citep[e.g.][]{gallazzi2005,panter2008,zahid2017,trussler2018}, integral field spectroscopic (IFS) surveys \citep[e.g. CALIFA, MaNGA, SAMI;][]{gonzalezdelgado2014,scott2017,lian2018c} and, at lower stellar masses, from stellar spectroscopy in local dwarf galaxies \citep[e.g.][]{kirby2013}.
In the majority of these cases, metallicities have been derived from rest-frame optical continuum observations.
Recently, \citet{zahid2017} presented a compilation of $Z_{\ast}$ estimates in the local Universe spanning $\simeq 7$ orders of magnitude in stellar mass (ranging from $M_{\ast} \simeq 10^4 \rm{M}_{\odot}$ to $M_{\ast} \simeq 10^{11}\rm{M}_{\odot}$) finding evidence for a continuous relation which rises from \logz \ $\simeq -2.5$ at the lowest stellar masses, up to \logz \ $\simeq 0.0$, flattening at stellar masses above $M_{\ast} \simeq 10^{10}\rm{M}_{\odot}$.
Determining whether or not a similar relation was already in place at earlier cosmic epochs is clearly of significant interest.

Unfortunately, estimates of $Z_{\ast}$ for galaxies at $z\gtrsim2$ are rare, primarily because the measurement requires a high signal-to-noise ratio (S/N) detection of the stellar continuum, an expensive and challenging observation for faint high-redshift sources.
Nevertheless, estimates have been made for small samples of lensed galaxies at $z \sim 2-3$ \citep{rix2004,quider,dessauges2010}, a number of unlensed sources at $z > 3$ \citep{sommariva2012} and a composite spectrum of star-forming galaxies at $z\simeq 2$ \citep{halliday2008}.
In all of these cases, the observations have been taken with ground-based optical spectrographs and therefore the stellar metallicities are based on rest-frame far-ultraviolet (FUV) spectra.
The distinction is important, because FUV-based metallicities are considered to be, to a first approximation, a measurement of the iron abundance in the photospheres of the young, massive, O- and B-type stars in the galaxy \citep[e.g.][]{halliday2008}.
This is not necessarily the case for optical-based stellar metallicities, which trace more evolved stars and sample longer star-formation timescales, and is certainly not the case for estimates of $Z_{\rm{g}}$ derived from optical nebular emission lines, which trace the young stellar population, but mainly the oxygen and nitrogen abundance.

Most commonly, stellar metallicities at high-redshift have been determined using a calibration of FUV photospheric absorption indices developed by \citet{rix2004}, and later extended in \citet{sommariva2012}.
However, the results from these studies have generally been inconclusive, especially with respect to the scaling relation between metallicity and stellar mass.
This has been, in part, a consequence the of small sample sizes and subsequent lack of dynamic range, but is also a consequence of the fact that these indices are also prone to significant contamination by interstellar medium (ISM) absorption \citep[e.g.][]{vidal-garcia2017}.
More recently, an alternative method for measuring $Z_{\ast}$, by fitting the full FUV spectrum at $\lambda \leq 2000 \mathrm{\AA}$, has been demonstrated by \citet{steidel2016} \citepalias{steidel2016} using the BPASSv2.0 stellar population synthesis models including massive stellar binaries \citep{eldridge2017}.
The method is a natural extension of the previous methods, making more complete and consistent use of the available data and should, in principle, place more robust constraints on $Z_{\ast}$ than the standard indices approach \citep[e.g.][]{walcher2011,conroy2018}.

Applying the full FUV spectral fitting method to a composite spectrum of 30 star-forming galaxies at $z=2.4$, \citet{steidel2016} found a low FUV-based stellar metallicity of $Z_{\ast}/Z_{\odot}\simeq0.1$, roughly a factor of five lower than the gas-phase metallicity derived from rest-frame optical nebular emission lines ($Z_{\rm{g}}/Z_{\odot}\simeq0.5$), which they interpreted as evidence for super-solar oxygen-to-iron ratios (O/Fe) in high-redshift star-forming galaxies.
This result was subsequently confirmed using a larger statistical sample of 150 star-forming galaxies at $z\sim2-3$ in \citet{strom2018} albeit via a slightly different method in which $Z_{\ast}$ and $Z_{\rm{g}}$ were simultaneously estimated from fitting optical nebular emission lines.
However, although \citet{steidel2016} established that star-forming galaxies at $z\gtrsim2$ have significantly sub-solar stellar metallicities (iron abundances), they did not explore the existence of a scaling relation of $Z_{\ast}$ with stellar mass.
Therefore, determining the stellar mass-metallicity relation at $z\gtrsim2$ remains a key open question in the study of galaxy evolution.

Fortunately, further progress can be made in this area using the deep spectroscopic data taken as part of the VANDELS survey \citep{mclure_vandels,pentericci_vandels}.
VANDELS is an ESO public spectroscopic survey providing exceptionally deep integration times (of up to 80 hours) for $\sim 2000$ galaxies at $z \gtrsim 1$.
At redshifts $2.5 < z < 5.0$, the VANDELS spectra cover the rest-frame FUV, allowing stellar metallicities to be estimated using the methods described above.
Moreover, since all of the VANDELS targets are drawn from the CDFS and UDS survey fields, accurate stellar masses can be determined for all sources using the deep multi-wavelength photometric catalogues available in these regions.
In this paper, we present a study of the stellar mass-metallicity relationship at $2.5 < z < 5.0$ using deep rest-frame FUV spectra from the VANDELS survey.

The structure of the paper is as follows.
In Section \ref{sec:data} we describe the selection of our VANDELS star-forming galaxy sample along with the other relevant datasets used in this work.
In Section \ref{sec:method} we begin by reviewing the metallicity information contained within rest-frame FUV galaxy spectra, and outline our method for determining FUV-based stellar metallicities.
In Section \ref{sec:results_stellar_mzr} we present our stellar mass-metallicity relation and compare to the predictions from three independent cosmological simulations.
In Section \ref{sec:discussion} we discuss some of the implications of our results, including an investigation into which physical parameters are potentially driving the observed stellar metallicities in our sample, and a discussion of O/Fe ratios in high-redshift star-forming galaxies.
Finally, in Section \ref{sec:conclusions} we summarize our results and conclusions.
Throughout this paper metallicities are quoted relative to the solar abundance taken from \citet{asplund2009} which has a bulk composition by mass of $Z_{\ast}=0.0142$, an iron mass fraction of $Z_{\ast,\rm{Fe}}=0.0013$ and an oxygen mass fraction of $Z_{\ast,\rm{O}}=0.0058$, and we assume the following cosmology: $\Omega_{M} =0.3$, $\Omega_\Lambda =0.7$, $H_0 =70$ km s$^{-1}$ Mpc$^{-1}$.
All magnitudes are quoted in the AB magnitude system \citep{oke1983}.

	\begin{figure}
        \centerline{\includegraphics[width=\columnwidth]{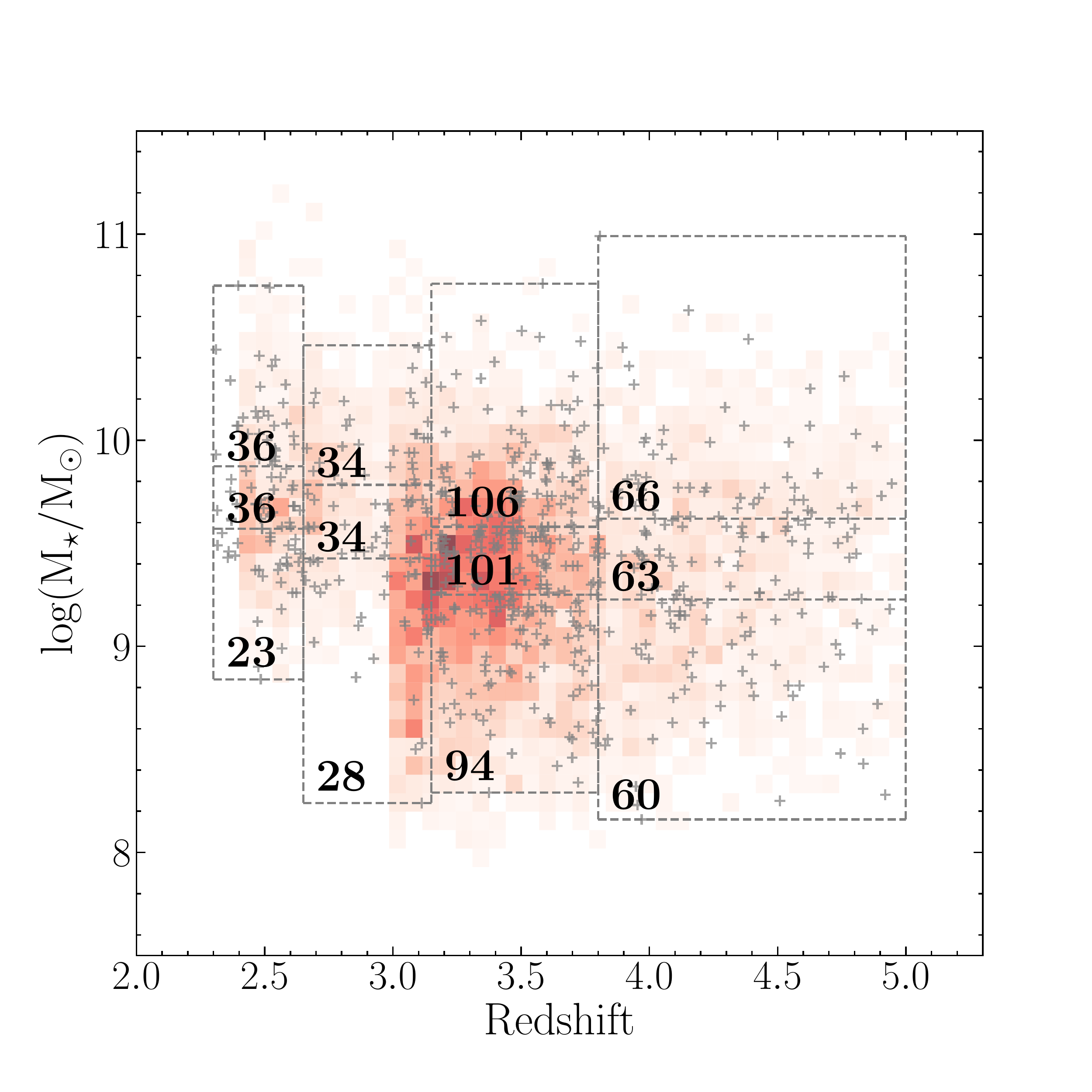}}
        \caption{Stellar mass versus redshift for the VANDELS DR2 spectra at $2.3 < z < 5.0$ showing (via the grey dashed grid) the bins in stellar mass and redshift used to produce composite spectra.
        The number within each box indicates the number of individual galaxies in that bin, which are marked by small grey `+' symbols.
        The background 2D histogram shows all the galaxies within the initial pre-selection catalogue with  $2.4 \leq z_{\rm{phot}} \leq 5.0$, illustrating how the observed galaxies are effectively randomly drawn from this population.}
        \label{fig:mass_redshift_dist}
    \end{figure}

\begin{table*}
    \centering
    \caption{Details of the VANDELS composite spectra stacked in bins of stellar mass and redshift.}\label{table:composite_info}
    \begin{tabular}{llrclrrc}
        \hline
        \hline
        Composite ID & \thead{Bin Range \\ \logm} & \thead{Median \\ \logm} & $\langle z \rangle$ & \thead{Wavelength \\ Coverage ($\rm{\AA}$)} & N$_{\rm{gal}}$ & \thead{Median S/N \\ per pixel} & log($Z_{\ast}$) \\
        \hline
        {\raggedleft $2.30 \leq z \leq 5.00$:} \\
        VANDELS-m1 & $\phantom{0}8.16\phantom{0} - \phantom{0}8.70$ & $8.51$ & $3.82$ & $1000 - 2000$ & $38$ & $6$ & $<-2.89$ \\
        VANDELS-m2 & $\phantom{0}8.70\phantom{0} - \phantom{0}9.20$ & $9.00$ & $3.67$ & $1000 - 2000$ & $131$ & $14$ & $-2.78\pm0.03$ \\
        VANDELS-m3 & $\phantom{0}9.20\phantom{0} - \phantom{0}9.50$ & $9.36$ & $3.47$ & $1000 - 2000$ & $153$ & $19$ & $-2.78\pm0.03$ \\
        VANDELS-m4 & $\phantom{0}9.50\phantom{0} - \phantom{0}9.65$ & $9.56$ & $3.50$ & $1000 - 2000$ & $111$ & $15$ & $-2.70\pm0.03$ \\
        VANDELS-m5 & $\phantom{0}9.65\phantom{0} - \phantom{0}9.80$ & $9.71$ & $3.48$ & $1000 - 2000$ & $95$ & $15$ & $-2.71\pm0.03$ \\
        VANDELS-m6 & $\phantom{0}9.80\phantom{0} - 10.00$ & $9.89$ & $3.34$ & $1000 - 2000$ & $74$ & $14$ & $-2.56\pm0.06$ \\
        VANDELS-m7 & $10.00\phantom{0} - 11.00$ & $10.24$ & $3.24$ & $1000 - 2000$ & $79$ & $12$ & $-2.42\pm0.06$ \\
        \hline
        {\raggedleft $2.30 \leq z < 2.65$:} \\
        VANDELS-z1-m1 & $\phantom{0}8.84\phantom{0} - \phantom{0}9.47$ & $9.30$ & $2.54$ & $1450 - 2000$ & $23$  & $13$ & $-2.64 \pm 0.06$ \\
        VANDELS-z1-m2 & $\phantom{0}9.49\phantom{0} - \phantom{0}9.81$ & $9.74$ & $2.50$ & $1450 - 2000$ & $36$  & $14$ & $-2.68 \pm, 0.05$ \\
        VANDELS-z1-m3 & $\phantom{0}9.82\phantom{0} - 10.75$ & $9.82$ & $2.50$ & $1450 - 2000$ & $36$  & $13$ & $-2.50 \pm 0.08$ \\
        \hline
        {\raggedleft$2.65 \leq z < 3.15$:} \\
        VANDELS-z2-m1 & $\phantom{0}8.24\phantom{0} - \phantom{0}9.39$ & $9.08$ & $2.98$ & $1200 - 2000$ & $28$  & $8$ & $-2.64 \pm 0.11$ \\
        VANDELS-z2-m2 & $\phantom{0}9.41\phantom{0} - \phantom{0}9.74$ & $9.57$ & $2.92$ & $1200 - 2000$ & $34$  & $9$ & $-2.63 \pm 0.10$ \\
        VANDELS-z2-m3 & $\phantom{0}9.78\phantom{0} - 10.46$ & $10.02$ & $2.97$ & $1200 - 2000$ & $34$  & $9$ & $-2.44 \pm 0.10$ \\
        \hline
        {\raggedleft$3.15 \leq z < 3.80$:} \\
        VANDELS-z3-m1 & $\phantom{0}8.29\phantom{0} - \phantom{0}9.22$ & $8.94$ & $3.50$ & $1100 - 2000$ & $94$  & $12$ & $-2.90 \pm 0.03$ \\
        VANDELS-z3-m2 & $\phantom{0}9.23\phantom{0} - \phantom{0}9.57$ & $9.43$ & $3.45$ & $1100 - 2000$ & $101$  & $16$ & $-2.82 \pm 0.03$ \\
        VANDELS-z3-m3 & $\phantom{0}9.58\phantom{0} - 10.76$ & $9.99$ & $3.50$ & $1100 - 2000$ & $106$  & $15$ & $-2.68 \pm 0.04$ \\
         \hline
        {\raggedleft$3.80 \leq z \leq 5.00$:} \\
        VANDELS-z4-m1 & $\phantom{0}8.16\phantom{0} - \phantom{0}9.20$ & $8.85$ & $4.31$ & \phantom{0}$900 - 2000$ & $60$  & $8$ & $-2.89 \pm 0.05$ \\
        VANDELS-z4-m2 & $\phantom{0}9.21\phantom{0} - \phantom{0}9.60$ & $9.41$ & $4.26$ & $\phantom{0}900 - 2000$ & $63$  & $9$ & $-2.71 \pm 0.11$ \\
        VANDELS-z4-m3 & $\phantom{0}9.61\phantom{0} - 11.00$ & $9.87$ & $4.26$ & \phantom{0}$900 - 2000$ & $66$  & $10$ & $-2.59 \pm 0.08$ \\
        \hline
    \end{tabular}
\end{table*}

\section{Data}\label{sec:data}

In this section we present the various observed and simulated datasets used in this work.
The dataset of primary interest, described in Section \ref{subsec:vandels}, is the rest-frame FUV spectra of star-forming galaxies at  $2.5 < z < 5.0$ provided by the VANDELS survey.
These high-redshift data are supplemented by a selection of FUV spectra of starburst galaxies and star-forming regions in the local Universe which are described in Section \ref{subsec:fos_ghrs_data}.
Finally, in Section \ref{subsec:simulation_description}, we discuss the data extracted from two state-of-the-art cosmological simulations \citep[\fiby and \simba;][]{paardekooper2015,dave2019} which are used to both test our method and to aid in the analysis and interpretation of our results.

\subsection{VANDELS}\label{subsec:vandels}

The primary spectroscopic data were obtained as part of the VANDELS ESO public spectroscopic survey \citep{mclure_vandels, pentericci_vandels}.
The VANDELS survey is a deep optical spectroscopic survey of the CANDELS CDFS and UDS fields with the VIMOS spectrograph on ESO's Very Large Telescope (VLT) targeting massive passive galaxies at $1.0 \leq z \leq 2.5$, bright star-forming galaxies at $2.4 \leq z \leq 5.5$ and fainter star-forming galaxies at $3.0 \leq z \leq 7.0$.
In this work, we focus exclusively on the star-forming galaxies at $z \leq 5$.

All galaxies were drawn from four independent $H$-band-selected catalogues.
The CDFS and UDS regions are covered by the CANDELS survey, and benefit from extensive WFC3/IR imaging \citep[CDFS-$HST$, UDS-$HST$;][]{grogin2011,koekemoer2011}.
In these fields, photometry catalogues produced by the CANDELS team were used \citep{galametz2013,guo2013}.
Within the wider-field regions, two bespoke PSF-homogenized catalogues were produced, primarily from publicly available ground-based imaging \citep[CDFS-GROUND, UDS-GROUND;][]{mclure_vandels}.
From these catalogues objects were pre-selected as potential spectroscopic targets based on robust photometric redshift estimates.
The bright star-forming galaxies were chosen to satisfy $2.4 \leq z_{\rm{phot}} \leq 5.0$ with an $i$-band magnitude of $i \leq 25$.
This $i$-band constraint was chosen to ensure the final spectra had sufficient signal-to-noise (S/N) to allow detailed analyses of individual objects.
The faint star-forming galaxies relevant to our study (i.e. in the redshift interval $3.0 \leq z_{\rm{phot}} \leq 5.5$) were selected to have $25 \leq H \leq 27 \land i \leq 27.5$ in the $HST$ regions \footnote{In the wider-field regions the limiting $i$-band magnitude is slightly brighter at $i \leq 26.0$.}.  
As discussed in \citet{mclure_vandels} (illustrated in their Fig. 3), this selection ensured that the sample of star-forming galaxies is consistent with being drawn from the main sequence of star-forming galaxies at all redshifts.
An additional constraint, enforced by the observing strategy, was placed on the $i$-band magnitudes so that slits were efficiently allocated to objects requiring 20, 40 and 80hr of integration time in a roughly 1:2:1 ratio \citep{mclure_vandels}.
The final spectroscopic sample of faint star-forming galaxies was an unbiased random (approximately 1 in 4) subsample of the pre-selection catalogue.
    
The observations and reduction of the VIMOS spectra are described in detail in the first data release paper \citep{pentericci_vandels}.
Briefly, observations were obtained with the ESO-VLT VIMOS spectrograph using the medium resolution grism which covers the wavelength range $4800 < \lambda_{\rm{obs}} < 10000 \rm{\AA}$ with a resolution of $R=580$ and a dispersion of 2.5 \rm{\AA} per pixel.
The median seeing across all observations was FWHM$=0.7$\asec.
The spectra were reduced using the \textsc{Easy-Life} pipeline \citep{garilli2012} which is an update of the algorithms originally described in \citet{scodeggio2005}.
\textsc{Easy-Life} performs all standard data reduction procedures including the removal of bad CCD pixels and sky subtraction at the individual exposure level.
The 2D spectrum for each individual exposure is extracted and resampled onto a common linear wavelength scale; these individual-exposure 2D spectra are then combined to create the final 2D spectrum for each object.
The 1D spectra are extracted, flux calibrated using the spectrophotometric standard stars and corrected for telluric absorption.
A final correction for atmospheric and galactic extinction is applied and the 1D spectra are normalized to the $i$-band photometry.
     
     \begin{figure*}
        \centerline{\includegraphics[width=7.5in]{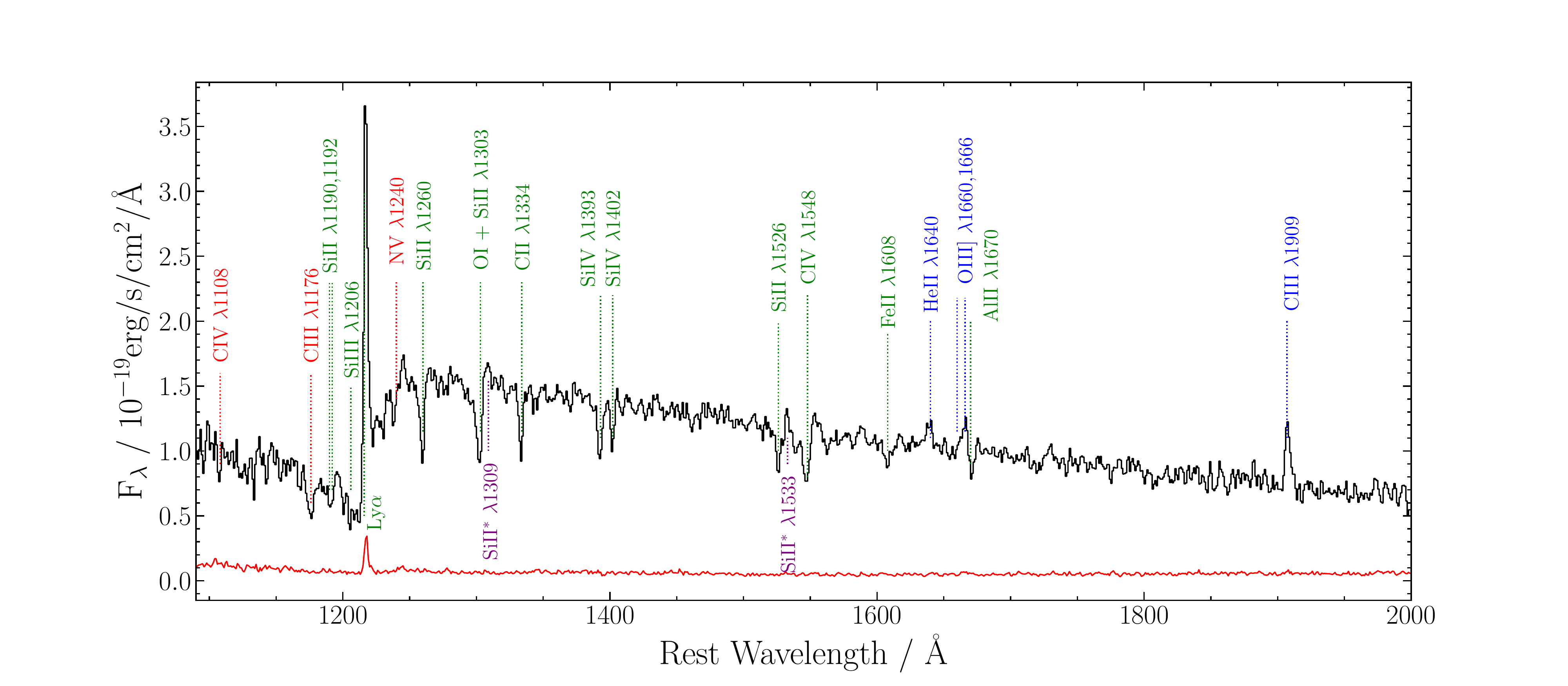}}
        \caption{An example composite spectrum (black curve) built from the spectra of the 131 galaxies in our sample within the mass range $8.7 <$ \logm \ $< 9.2$ (VANDELS-m2; see Table \ref{table:composite_info} for details) along with the error spectrum (red line) estimated via a bootstrap re-sampling technique.
        The galaxies contributing to the composite span the full redshift range of our sample ($2.3 < z < 5.0$) with a mean redshift of $z=3.67$.
        To account for redshift differences, the flux of each individual spectrum was scaled to the flux that would be observed at the mean redshift.
        The labels indicate prominent emission/absorption features color-coded by their physical origin: interstellar absorption (green), stellar absorption (red), nebular/stellar emission (blue) and fine structure emission (purple).
        The remainder of the FUV spectrum is dominated by absorption due to heavy photospheric line-blanketing in the atmosphere of massive O- and B-type stars; these are the regions that were used to constrain the stellar metallicity (Section \ref{subsec:statistical_method} for details).}
        \label{fig:full_stacked_spectrum}
    \end{figure*}

We draw our sample from the second VANDELS data release\footnote{Data available through the consortium database at the link http://vandels.inaf.it/db} which comprises roughly $65\%$ $(1362/2106)$ of the full survey.
Redshifts for all of the spectra have been determined by members of the VANDELS team and assigned a redshift quality flag as described in \citet{pentericci_vandels}.
We only consider galaxies at $2.3 \leq z_{\rm{spec}} \leq 5.0$ and with a redshift quality flag 3 or 4 (corresponding to a $\geq95\%$ probability of being correct), leaving a total of 681 galaxies.
Fig. \ref{fig:mass_redshift_dist} shows the distribution of our final sample in the mass-redshift plane compared to the underlying distribution of all $2.4 \leq z_{\rm{phot}} \leq 5.0$ star-forming galaxies in the pre-selection catalogue.
We have extended our selection to $z_{\rm{spec}} \geq 2.3$ so that we include galaxies whose photometric redshift was slightly overestimated at the lower redshift boundary of the star-forming galaxy selection ($z_{\rm{phot}}=2.4$).
The upper redshift limit of $z=5$ is enforced primarily because above this redshift the FUV wavelength coverage becomes prohibitively small.
The average redshift of our final sample is $\langle z \rangle = 3.5 \pm 0.6$.
Stellar masses for the full VANDELS sample were derived from SED fitting using \citet{bruzual2003} templates with solar metallicity and assuming exponentially declining star-formation histories as described in \citet{mclure_vandels}.
We also derived stellar masses for our $2.5 < z < 5.0$ sample using \textsc{fast++}, a rewrite of \textsc{fast} \citep{kriek2009} described in \citet{schreiber2018}.
We adopted the same set of parameters as \citet{mclure_vandels}, but allowed for rising star-formation histories using a delayed exponentially declining model ($M_{\ast} \propto te^{t/\tau}$) and a range of metallicities values ($0.3-2.5 \times \mathrm{Z}_{\odot}$).
The stellar masses returned from these two procedures were fully consistent, but for this work we adopt the \textsc{fast++} values.

As discussed in \citet{mclure_vandels}, the pre-selection catalogue contains star-forming galaxies consistent with being drawn from the star-forming main sequence across the full redshift range.
Therefore, since the final selection is a random subsample of galaxies from the pre-selection catalogue, the galaxies in our final sample can be considered typical star-forming galaxies for their stellar mass.
The stellar mass range of the final sample is $8.2 <$ \logm $<11.0$ with a median specific star-formation rate of sSFR$=4.4$ \Gyr$^{-1}$.

\subsubsection{Composite FUV spectra}

The typical continuum S/N is too low to extract metallicity information, therefore we focus our analysis on stacked spectra in bins of stellar mass and redshift.
The mass and redshift bins are illustrated in Fig. \ref{fig:mass_redshift_dist} and detailed in Table \ref{table:composite_info}.
In total we chose four redshift bins which were selected to encompass roughly equivalent intervals of cosmic time ($\approx 400$ Myr), with three stellar-mass bins per redshift.
We also analyse seven composite spectra stacked in bins of stellar mass across the full redshift range, with the stellar mass ranges given in Table \ref{table:composite_info}.
The mass bins were chosen manually with the aim of keeping the bin widths as narrow as possible whilst also ensuring that they contained enough galaxies to have an acceptable S/N.

The final stacked composite spectra were formed by first shifting each contributing spectrum into the rest-frame using $z_{\rm{spec}}$. 
To correct for redshift differences, the flux of each spectrum was scaled to the flux that would be observed at the mean redshift of the stack.
For the stellar-mass redshift stacks these flux-correction factors are all $\lesssim 5 \%$; for the stellar-mass only stacks the maximum average correction is $15\pm7\%$ which occurs in the highest redshift bin.
We have confirmed that our results do not change significantly if, instead of this flux-correction method, the spectra are simply normalized and then combined.
The primary benefit of preserving the absolute flux values is that an error spectrum can be robustly estimated.
The individual spectra were then resampled onto a common wavelength grid, which varied slightly depending on the redshift range (see Table \ref{table:composite_info}), with a dispersion of $1 \rm{\AA}$ per pixel. 
The final flux at each dispersion point was taken as the median of all the individual flux values after rejecting $3\sigma$ outliers, and the $1\sigma$ error was calculated from bootstrap re-sampling of the individual flux values.
An example of one composite spectrum is shown in Fig. \ref{fig:full_stacked_spectrum}.
The effective spectral resolution element of the composites is $3.0 \rm{\AA}$.

\subsection{FOS-GHRS Local Sample}\label{subsec:fos_ghrs_data}

The main high-redshift sample was supplemented by a sample of local galaxies for which published rest-frame FUV spectra are available.
The motivation for including a local reference sample was that the majority of published stellar metallicities in the local Universe are derived from rest-frame optical spectra \citep[e.g.][]{gallazzi2005,panter2008,gonzalezdelgado2014,zahid2017}, and it is not clear that optical and FUV light will originate from the same population of stars or trace the same abundance type.
Even in the cases where rest-frame UV spectra are analysed, the focus is primarily on deriving gas-phase metallicities (not stellar metallicities) from the depths of ISM absorption features \citep[e.g.][]{leitherer2011, zetterlund2015,faisst2016}.
To our knowledge, there are no local examples of stellar metallicities derived solely from global FUV continuum fitting.
For this reason, we have constructed a comparison sample of local starbursts and star-forming galaxies from the FUV spectroscopic atlas presented in \citet{leitherer2011} which is comprised of spectra taken with the Faint Object Spectrograph (FOS) and Goddard High Resolution Spectrograph (GHRS).

Our sample is drawn from the 46 rest-frame UV spectra observed with the Faint Object Spectrograph (FOS) and the Goddard High Resolution Spectrograph (GHRS) on-board the \emph{HST} presented in \citet{leitherer2011}.
These 46 spectra are drawn from 28 individual galaxies (i.e. some spectra are just different regions of the same galaxy).
The spectral resolution of the individual spectra spans the range $0.5-3\rm{\AA}$ depending on the instrument, aperture size and physical extent of the object being observed.
For consistency with the the VANDELS data, we smooth all spectra to a common $3\rm{\AA}$ resolution.
We also require that the spectra have wavelength coverage in the interval $1410 \leq \lambda \leq 1450 \rm{\AA}$ to enable an analysis of normalized composite spectra, and finally that the corresponding galaxy has a measurement of absolute $K$-band magnitude that we can use to estimate the stellar mass using the relation from \citet{mcgaugh2014}.
This selection leaves 26/46 of the original spectra from 18/28 of the original local starbursts and star-forming galaxies presented in \citet{leitherer2011}.
A list of the individual spectra used in this work is given in the Appendix (Table \ref{table:fos_ghrs_sample_appdx}). 
Composite spectra were constructed in three bins of stellar mass as outlined in Table \ref{table:fos_ghrs_sample_appdx}.
To create the composites, all individual spectra were median combined on the same wavelength grid used for the VANDELS galaxies.
In this case, an error spectrum was estimated by propagating the error spectra of each of the individual spectra (bootstrapping could not be applied in this case since the individual spectra were normalized).

\subsection{Simulation Data}\label{subsec:simulation_description}

To compare the observations to simulation predictions we have extracted data from two state-of-the-art cosmological hydrodynamical simulations: \fiby \ \citep[e.g][]{johnson2013,paardekooper2015,cullen2017} and \simba \ \citep{dave2019}.
In Section \ref{subsec:systematic_uncertainties} we use the simulation data to assess the accuracy our adopted method for measuring stellar metallicities and in Section \ref{sec:results_stellar_mzr} we compare the observed stellar mass-metallicity relationship to the simulation predictions.
Having two independent simulations allows us to account for systematic variations between their methodologies and theoretical predictions.
Here, we give a brief overview of the simulation details but refer the reader to the references above for further information.

\subsubsection{\textsc{FiBY}}

The \fiby \ simulation suite is a set of high-resolution cosmological hydrodynamical simulations using a modified version of the GADGET code used in the Overwhelmingly Large Simulations (OWLS) project \citep{schaye2010}. 
The code tracks metal pollution for 11 elements: H, He, C, N, O, Ne, Mg, Si, S, Ca and Fe and calculates the cooling of gas based on line-cooling in photoionization equilibrium for these elements \citep{wiersma2009} using tables pre-calculated with CLOUDY v07.02 \citep{ferland1998}. 
Furthermore, the simulation incorporates full non-equilibrium primordial chemistry networks \citep{abel1997,galli1998,yoshida2006} including molecular cooling functions for H$_2$ and HD. 
Star formation is modelled using the pressure law implementation of \citet{schaye2008}, which yields results consistent with the Schmidt-Kennicutt law \citep{schmidt1959, kennicutt1998_schmidt}. 
The simulations include feedback from stars by injecting thermal energy into the neighboring particles \citep{dalla_vecchia2012}. 
Element yields from Type Ia and core-collapse supernova (CCSNe) are implemented following the prescription of \citet{wiersma2009}.
In this work we will focus on the \fiby$\_$\textsc{XL} simulation which has individual gas and star particle masses of \logm \ $= 5.68$ and covers a co-moving volume of $({32 \: \mathrm{Mpc})^3}$.
The lowest redshift simulated in the \fiby \ simulation is $z=4$, and this is the closest redshift to the mean redshift of our sample ($\langle z \rangle=3.5$).
Therefore, for our \fiby \ comparison sample, we extracted the 591 galaxies in \fiby$\_$\textsc{XL} with \logm \ $> 8.0$ at $z=4$.
The maximum galaxy stellar mass is $M_{\ast}=3.8 \times 10^{10}$ M$_{\odot}$, with a median value of $4.0 \times 10^8$ M$_{\odot}$.

\subsubsection{\textsc{Simba}}

The \simba \ simulation suite is based on the \textsc{GIZMO} cosmological gravity plus hydrodynamics solver \citep{hopkins2015,hopkins2017} and is described in detail in \citet{dave2019}.
The code implements a H$_2$-based star formation rate, with H$_2$ fractions calculated using the sub-grid model of \citet{krumholz2011}.
The chemical enrichment model tracks the same 11 elements as in \fiby, with radiative cooling and photoionization heating modeled using the \textsc{Grackle-3.1} library \citep{smith2017}, and element yields calculated for CCSNe, Type-Ia SNe and Asymptotic Giant Branch stars following the prescriptions of \citet{nomoto2006}, \citet{iwamoto1999} and \citet{oppenheimer2008}, respectively.
\simba \ employs a mass outflow rate scaling with galaxy stellar mass motivated by the results derived from the extremely high resolution FIRE zoom simulations \citep{muratov2015,angles-alcazer2017}.
In this work we focus on the data from the m50n1024 simulation which has a co-moving box length of 50 $h^{-1}$Mpc and individual gas/star element resolution of $2.28\times 10^6$ M$_{\odot}$.
For consistency with \fiby, we focus on the $z=4$ snapshot in the simulation.
From this, we extract a sample of 1749 galaxies down to a minimum stellar mass of $4.5 \times 10^8$ M$_{\odot}$, with a maximum stellar mass of $1.7 \times 10^{11}$ M$_{\odot}$ and a median value of $8.2 \times 10^8$ M$_{\odot}$.

    \begin{figure}
        \centerline{\includegraphics[width=\columnwidth]{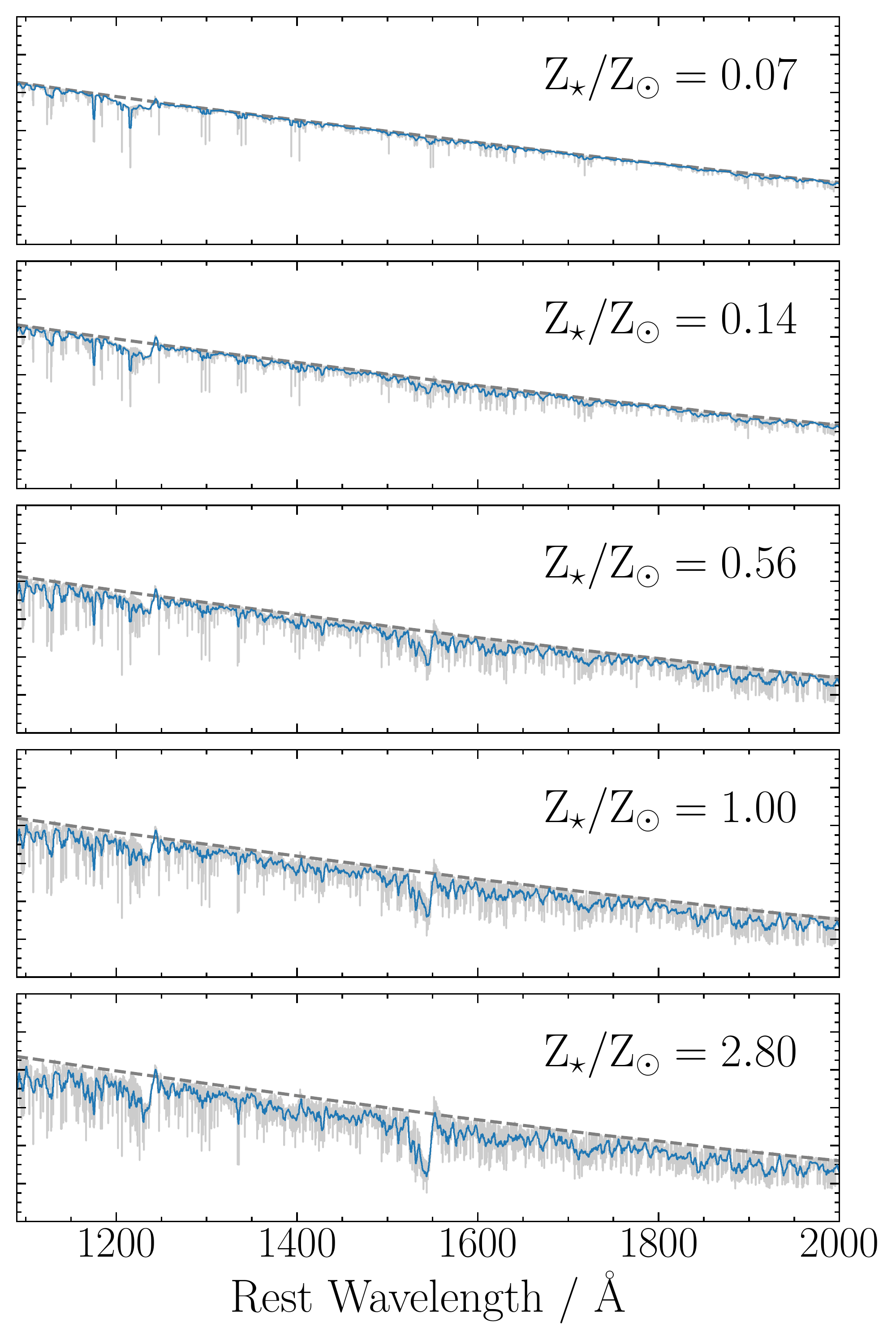}}
        \caption{The Starburst99 (SB99) high-resolution WM-Basic stellar population models used in this work.
        All models assume a constant star-formation rate over 100 \Myr.
        Each panel shows the model FUV spectra at one of the five default metallicities provided by the Starburst99 Geneva tacks ($Z_{\ast}=0.001, 0.002, 0.008, 0.014, 0.040$) running from the lowest metallicity (top panel) to the highest (bottom panel).
        Values of $Z_{\ast}/\mathrm{Z}_{\odot}$ are indicated in the legend.
        The background grey spectrum shows the full resolution SB99 models (0.4 $\rm{\AA}$) and the blue spectrum shows the models at the resolution of the VANDELS data ($3 \rm{\AA}$). 
        The grey dashed line shows the intrinsic stellar emission (i.e. before passing through the stellar photosphere).
        The figure illustrates how increasing the photospheric metallicity results in heavier FUV line-blanketing which, at the resolution of our data, results in a depression of the FUV continuum at all wavelengths.}
        \label{fig:wmbasic_sb99_spectra}
    \end{figure}

\section{Measuring Stellar Metallicities}\label{sec:method}

We now describe our method for estimating stellar metallicities from rest-frame FUV spectra.
In Section \ref{subsec:z_indicators} we describe the metallicity information content of the rest-frame FUV spectra of star-forming galaxies.
The stellar population synthesis models we use to compare to the data are discussed in Section \ref{subsec:sps_models}.
Then, in Section \ref{subsec:statistical_method}, we describe the statistical method we employ to robustly constrain the metallicity.
Finally, in Section \ref{subsec:systematic_uncertainties} we discuss various tests of our method. 

\subsection{Stellar Metallicity Indicators in the UV}\label{subsec:z_indicators}

To begin with, it is worthwhile reviewing what metallicity information is contained within the FUV spectrum of a star-forming galaxy.
Figs. \ref{fig:full_stacked_spectrum} and \ref{fig:wmbasic_sb99_spectra} serve as a useful references for this discussion.
From simple photon energy arguments, we know that the rest-frame FUV is dominated by the light from young, massive, O- and B-type stars.
These stars emit a continuum which contains absorption (and potentially emission) features due to elements within the stellar photosphere and the expanding stellar wind. 
The strength of these features is naturally a strong function of the total photospheric metallicity \citep[e.g.][]{leitherer2010}.

The most prominent absorption features are typically the strong stellar wind lines produced from the stellar atmospheres, namely \nv, \siiv \ and \civ.
These lines are often extremely broad, blueshifted by ${\sim 1000 \: \mathrm{kms}^{-1}}$ and exhibit P-Cygni line profiles. 
They are useful metallicity diagnostics since the strength of the stellar winds, and hence the line profiles, have a strong metallicity dependence which is linked to the metallicity-dependent mass-loss rates \citep{kudritzki2000,puls2008}.
Naturally, however, these lines are also sensitive to the IMF and star-formation history.

Outside these strong wind features, the rest-frame FUV shortward of $\lambda=2000\rm{\AA}$ contains an abundance of stellar photospheric absorption features, primarily due to transitions of highly ionized iron \citep{dean1985, brandt1998}.
The individual absorption features have low equivalent widths which require high spectral resolution and S/N to be seen individually, but the resulting strong photospheric line blanketing can be seen in unresolved O- and B-star populations in local starburst galaxies, star-forming regions and even high-redshift galaxies \citep{leitherer2011,halliday2008}.

    \begin{figure*}
        \centerline{\includegraphics[width=7.5in]{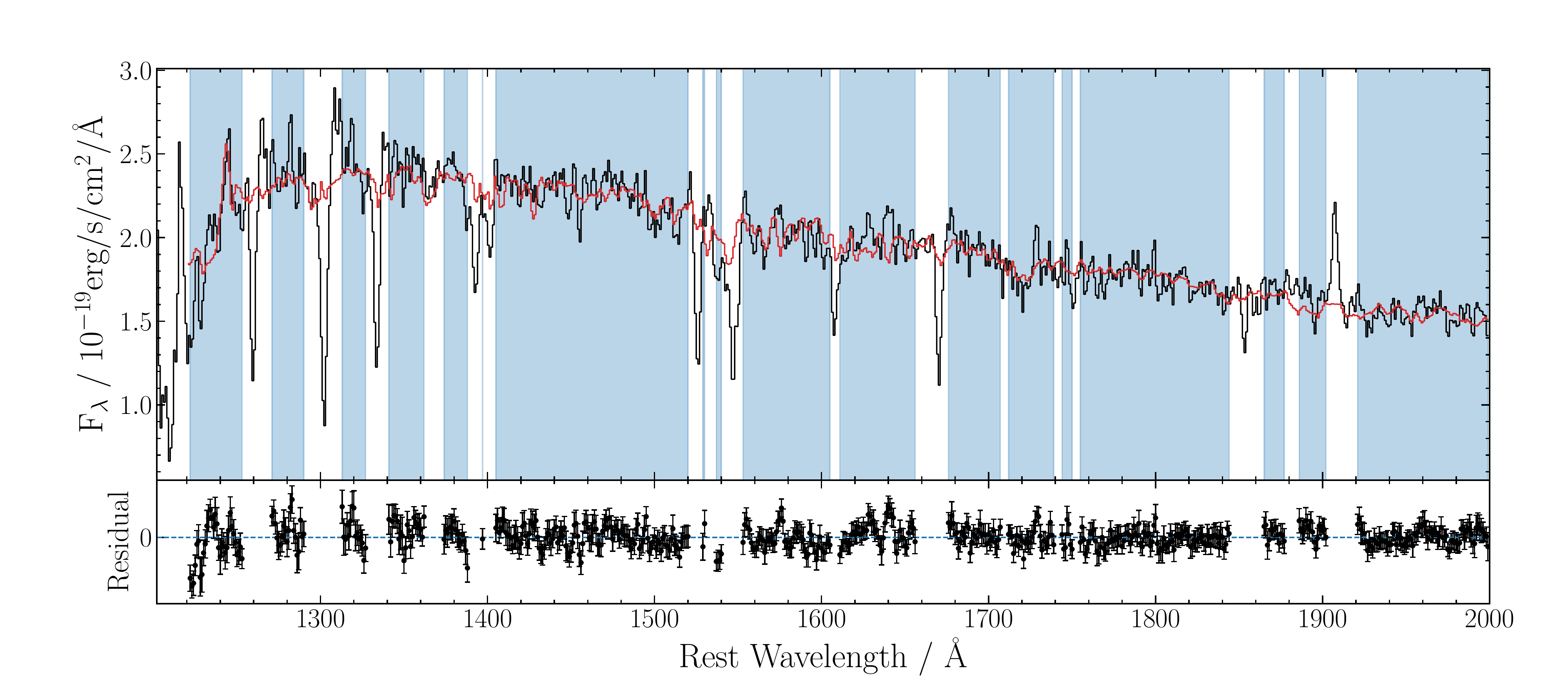}}
        \caption{An example of a full spectral fit to the VANDELS-m5 spectrum (see Table \ref{table:composite_info} for details).
        The upper panel shows the observed composite spectrum in black with the wavelength regions used in the fitting shaded in blue.
        The best-fitting Starburst99 WM-Basic spectrum is over-plotted in red.
        The lower panel shows the fit residuals; the reduced chi-squared value for this particular fit is $\chi^2_r=1.03$.}
        \label{fig:example_fit}
    \end{figure*}

These photospheric absorption features have been utilized in the past as metallicity indicators, for example \citet{rix2004} developed the `1978 index' at $1935-2050\rm{\AA}$, a wavelength regime which is believed to be dominated by \feiii \ transitions in early B-type stars. 
The 1978 index, which is thought to be a direct probe of the iron abundance ([Fe/H]), was used to derive the stellar metallicity of a composite spectrum of 75 star-forming galaxies at $z\approx2$ by \citet{halliday2008}.
Additional wavelength regions (e.g. \fev, \sIiii, \ciii, \sv) have been investigated by other authors and shown to be similarly sensitive to the photospheric abundance \citep[e.g][]{sommariva2012}.
These individual absorption features are analogous to the Lick index system developed for constraining abundances from rest-frame optical spectra \citep[][]{faber1985,thomas2003}.
However, \citet{vidal-garcia2017} have recently demonstrated that this approach is limited by the fact that the many of these indices are significantly contaminated by ISM absorption lines.

An alternative approach, demonstrated most recently by \citet{steidel2016}, is to simply fit the full FUV spectra directly, thereby constraining the FUV abundance by using all of the metallicity-sensitive regions simultaneously.
Indeed, similar approaches have proven successful for constraining stellar metallicities at optical wavelengths \cite[e.g.][]{panter2008,gonzalezdelgado2014}.
The obvious merit of this approach is that potentially relevant information is not being thrown away \citep[which may be the case when focusing on specific, narrow, wavelength regions, see e.g.][]{conroy2018}.
On the other hand, the fact that the absorption features across the full FUV spectrum are a result of a number of different element species means there is some ambiguity as to what abundance is actually being measured.
\citet{steidel2016} (among others) have pointed out that, because the vast majority of the FUV line-blanketing is a result of transitions in highly-ionized iron, a metallicity derived by fitting the FUV spectrum is, to a first approximation, a measure of the iron abundance in the stellar photosphere.
The point is demonstrated in Table 1 of \citet{leitherer2011}, which provides a comprehensive list of the photospheric spectral lines identified in star-forming galaxies both locally and from lensed high-redshift sources \citep{pettini2000}.
In this work we will adopt a full spectra fitting approach similar to \citet{steidel2016}, which is outlined in detail in Section \ref{subsec:statistical_method}.
Throughout this paper we refer to the stellar metallicity derived from rest-frame FUV observations as $Z_{\ast}/Z_{\odot}$ (or \logz), but emphasize that this should be understood as a proxy for the iron abundance. 

A final important point to note is that, while these FUV absorption features are strongly dependent on metallicity, the low-order shape of the FUV continuum is not (in contrast to the case at optical wavelengths).
By far the dominant factor in determining the FUV continuum shape is the wavelength-dependent dust attenuation law.
Interestingly, the attenuation law derived from local starbursts by \citet{calzetti2000} still appears to be a good approximation for the average shape of the attenuation law for star-forming galaxies at high-redshift with \logm \ $\gtrsim 9.5$ \citep[e.g.][]{cullen2017,cullen2018,mclure2017} , although this is still a matter of debate \citep[e.g.][]{reddy2018}, and object by object variation, and even some stellar mass dependence, is expected \citep[e.g.][]{kriek2013}.
Crucially, however, the lack of metallicity dependence on the global continuum shape (i.e the overall curvature of the FUV spectrum) negates any strong degeneracies between the metallicity and the wavelength-dependence/normalisation of the dust attenuation.

\subsection{Stellar Population Synthesis Models}\label{subsec:sps_models}

To derive metallicities from the observed FUV spectra we require models to compare to the data.
A number of different stellar population synthesis models exist \citep[e.g.][]{bruzual2003,molla2009,leitherer2010,eldridge2017}, but in this work  we have opted to use the Starburst99 (SB99) high-resolution WM-Basic theoretical stellar library described in \citet{leitherer2010}.
Our choice was motivated partly for clarity, and also by the fact that the WM-Basic SB99 spectra are the highest resolution FUV spectral models available, and have been extensively tested against individual spectra of hot stars in the Galaxy and Magellanic Clouds.
We argue that the WM-Basic SB99 models still represent the most robust stellar population models available for comparing to spectroscopic data in the FUV.
However, the SB99 models do not account for some physical effects that are known to be important in massive star evolution, such as the prevalence of binary stars and the phenomenon of `quasi-homogeneous evolution' at low metallicities \citep[e.g.][]{yoon2005,eldridge2012}.
In Appendix \ref{sec:bpassv2_comp} we have demonstrated that using stellar populations models which do account for these phenomena \citep[e.g. BPASSv2.1;][]{eldridge2017} would not affect our main results.

We considered constant star-formation rate models from the latest version of SB99 \citep{leitherer2014}, assuming the weaker-wind Geneva tracks with stellar rotation and single-star evolution.
We considered the Geneva evolution tracks at $Z_{\ast}=(0.001, 0.002, 0.008, 0.014, 0.040)$, and an $\eta=2.3$ IMF slope in the mass range \mbox{$0.5 < M_{\ast}/\mathrm{M}_{\odot} < 100$} corresponding to the default \citet{kroupa2001} IMF.
The high-resolution WM-basic spectra provided by Starbust99 cover the wavelength region $900-3000 \rm{\AA}$ at a dispersion of $0.4 \rm{\AA}$/pixel, so we smoothed and resampled the models to match the resolution ($\approx 3 \rm{\AA}$) and sampling ($1 \rm{\AA}$/pixel) of the VANDELS composites (Fig. \ref{fig:wmbasic_sb99_spectra}).
We considered constant star-formation models over timescales of 100, 300 and 500 \Myr \ but adopt the 100 \Myr \ models as our fiducial set (we discuss the motivation for this choice below).

Nebular continuum emission was added to all models using \textsc{cloudy v17.00} \citep{ferland2017} following the method described in \citet{cullen2017}, assuming \hii \ region parameters consistent with the current best estimates at $z\simeq2-3$ \citep{strom2018}.
When fitting the observed data we considered models with both maximal nebular contribution ($f_{\mathrm{esc}}=0\%$) and zero nebular contribution ($f_{\mathrm{esc}}=100\%$).
However, since the average escape fraction is measured to be relatively low in high-redshift star-forming galaxies \citep[$\lesssim 20\%$;][]{grazian2017,fletcher2018,steidel2018}, we adopt the models including maximal nebular continuum as our fiducial set.
We also note that, since the nebular continuum only acts to alter the shape of the continuum it is degenerate with the dust prescription, not with the stellar metallicity.
Assuming $f_{\mathrm{esc}}=100\%$ does not change the results presented here.

Finally, in Section \ref{subsec:systematic_uncertainties}, we will discuss how we tested our FUV-fitting method using synthetic FUV spectra derived from \fiby \ and \simba \ simulation data.
For generating these synthetic spectra we also used WM-Basic SB99 models, but in this case considered the instantaneous burst models generated at $1000$ time-steps split logarithmically between $1$ \Myr \ and $1.5$ \Gyr.
These instantaneous burst models are required so that the synthetic spectra can be constructed to match the known star-formation histories and chemical-abundance histories of the simulations.
The instantaneous burst models assume a total stellar mass within the burst of $10^6$ M$_{\odot}$ with all other parameters (e.g. IMF, metallicity range) being identical to the constant star-formation rate models.
As will be described in more detail below, the synthetic spectra were constructed from these instantaneous burst models using the average star-formation and chemical-abundance histories of the simulated galaxies following the method outlined in \citet{cullen2017}.

\subsection{A statistical estimate of the stellar metallicity}\label{subsec:statistical_method}

To fit the SB99 models to our data we adopted a Bayesian forward-modeling approach.
The method follows the familiar approach of model fitting using Bayes' theorem with,
\begin{equation}
P(\theta|D)\propto P(D|\theta)P(\theta),
\end{equation}
where $P(\theta|D)$ is the posterior probability on the input parameters $\theta$ given the data (D), $P(D|\theta)$ is the likelihood function ($L$), and $P(\theta)$ is the prior. 
Assuming the error bars are Gaussian and independent, the logarithm of the likelihood function is given by,
\begin{equation}
\mathrm{ln}(L) = -\frac{1}{2}\sum_{i}\bigg[\frac{(f_{i}-f(\theta)_{i})^2}{\sigma_{i}^2} + \mathrm{ln}(2\pi\sigma_{i}^2)\bigg]
\end{equation}
where $f$ is the observed flux, $f(\theta)$ is the model flux for a given set of parameters $\theta$, and $\sigma$ is the error on the observed flux.
The summation is over all wavelength pixels included in the fit.
For our model we adopted four free parameters: the logarithm of the stellar metallicity (log(Z$_{\ast}$/Z$_{\odot}$)) and three parameters used to fit the overall continuum shape, based on a physically motivated parameterization of the dust attenuation law.

    \begin{figure*}
        \centerline{\includegraphics[width=6.5in]{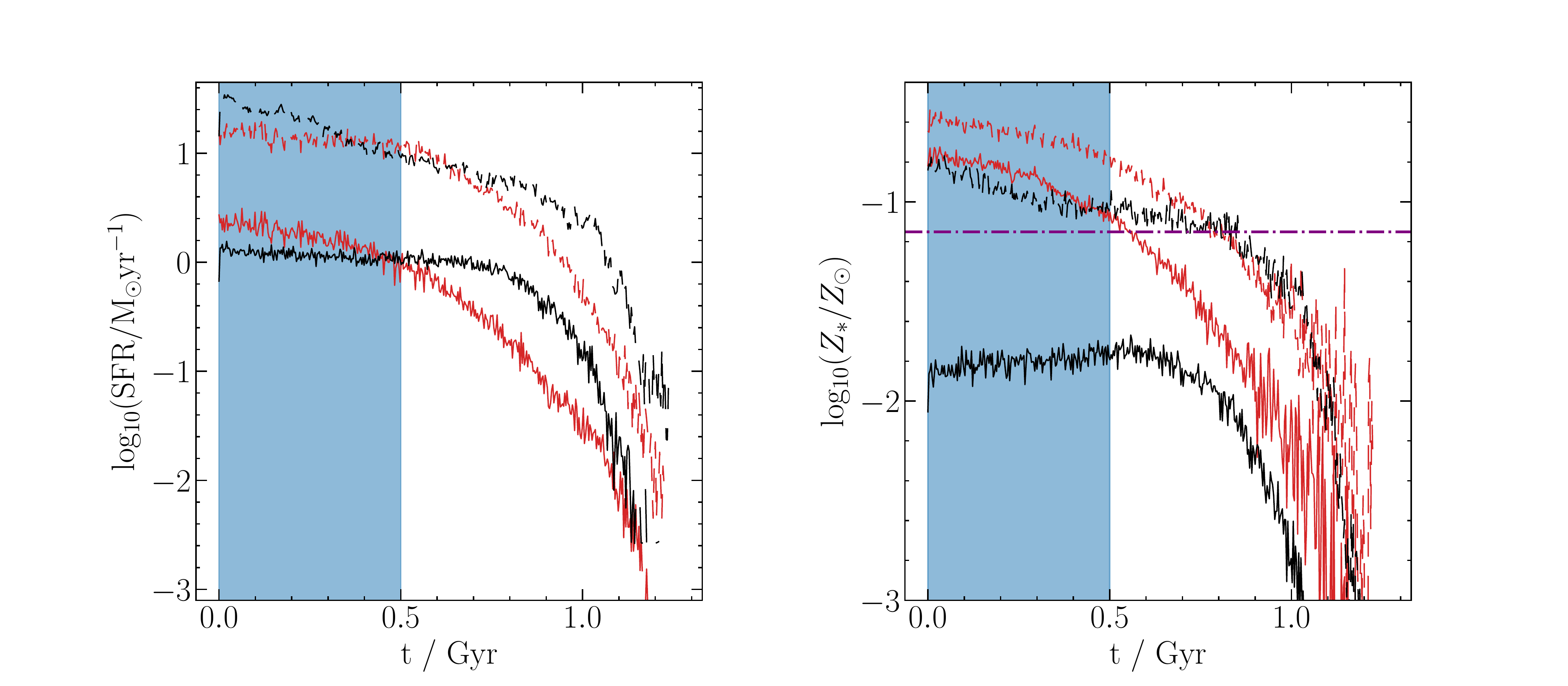}}
        \caption{The star-formation history (left-hand panel) and chemical-abundance history (in our case iron abundance; right-hand panel) of galaxies from the \fiby \ (red curves) and \simba \ (black curves) simulations at $z=4$.
        The values on the x-axis are lookback times relative to $z=4$.
        In each panel the solid curves show the average histories for galaxies with $8.5 <$ \logm \ $< 9.5$ and the dashed curves show galaxies with \logm \ $> 9.5$, roughly representative of the range of stellar masses in our sample.
        Despite the obvious differences, all curves show a rapid increase in star formation rate and metallicity over the first $\sim$ \Gyr \ follow by a more gradual evolution over the final $\sim 500$ \Myr \ (blue shaded region).
        The purple dot-dashed line in the right-hand panel indicates the lower metallicity limit of the SB99 WM-Basic models ($Z_{\ast}/Z_{\odot}=0.07$).}
        \label{fig:sim_sfr_chem}
    \end{figure*}

The dust law parameterization, taken from \citet{salim2018}, is a modification of the \citet{calzetti2000} attenuation law for starburst galaxies which allows for the slope of the curve to be modified and includes a prescription for the UV bump at $2175\rm{\AA}$. 
It is given by:
\begin{equation}
A_{\lambda,\mathrm{mod}}=\frac{A_{V}}{R_{V,\mathrm{mod}}}\bigg[k_{\lambda,\mathrm{Calz}}\frac{R_{V,\mathrm{mod}}}{R_{V,\mathrm{Calz}}}\bigg(\frac{\lambda}{5500\rm{\AA}}\bigg)^{\delta}+D_{\lambda}\bigg],
\end{equation}
where $A_{V}$ is the absolute $V$-band attenuation (i.e. the normalization of the attenuation law), $k_{\lambda,\mathrm{Calz}}$ is the total-to-selective attenuation curve for the \citet{calzetti2000} law and $\delta$ is the power law exponent used to modify the slope of the \citet{calzetti2000} law.
$R_{V,\rm{mod}}$ is the modified total-to-selective attenuation ratio which is simply a function of $\delta$ \citep[see equation 4 in][]{salim2018} and $R_{V,\rm{Calz}}=4.05$.{}
Finally, $D(\lambda)$ is the Drude profile which is commonly used as the functional form of the UV bump at $2175\rm{\AA}$ and is given by:
\begin{equation}
D_{\lambda}(B)=\frac{B\lambda^2\Delta^2}{[\lambda^2-\lambda_c^2]^2-\lambda^2\Delta^2}
\end{equation}
where $\lambda_c$ is the central wavelength of the feature ($2175\rm{\AA}$), $\Delta$ is the width ($350\rm{\AA}$) which are both held fixed, and B is the amplitude, which we allowed to vary\footnote{Although the peak of the UV bump at $2175\rm{\AA}.$ is outside the range of our fitting it can still potentially affect the shape of the UV continuum due to its width ($350\rm{\AA}$)}.
This dust attenuation prescription therefore has three free parameters: $A_{V}$ (normalisation), $\delta$ (slope) and $B$ (UV bump strength).
When $\delta=0$ and $B=0$ the parameterization simply becomes the \citet{calzetti2000} law.
We note that the role of the three dust parameters is to fit the shape of the continuum, which, as discussed above, has little dependence on the stellar metallicity (Fig. \ref{fig:wmbasic_sb99_spectra}). 

To perform the fitting, we used the nested sampling code \textsc{multinest} \citep{feroz2008,feroz2009}\footnote{We accessed \textsc{multinest} via the python interface \textsc{pymultinest} \citep{buchner2014}.}, which is an implementation of the nested sampling algorithm described in \citet{skilling2006}.
Nested sampling is an alternative to the traditional Markov Chain Monte Carlo (MCMC) method of sampling the posterior distribution for a given Bayesian inference problem, which enables the extraction of 1D posterior distributions for any given parameter in the model by marginalizing over all other free parameters.
We adopted simple flat priors on all the model parameters.
The prior in \logz \ is imposed by the SB99 models to be $-1.15 <$ \logz \ $< 1.45$. 
As the models are only provided at five fixed metallicity values, we linearly interpolated the logarithmic flux values between the models to generate a model at any metallicity within the prescribed range.
We experimented with other interpolation schemes (e.g. interpolating in flux linear flux values) but found that this did not strongly affect our results, and performed simple tests (e.g. recovering the known metallicity of the Starburst99 templates) to ensure the interpolation scheme was not strongly biasing the recovered metallicities.
For $A_{\rm{V}}$ and $B$ we considered values in the range $0 - 5$ and for $\delta$ we considered $-1.0 < \delta < 1.0$.

A final point to note is that not all wavelength pixels were included in the fitting.
As discussed above (and illustrated in Fig. \ref{fig:full_stacked_spectrum}) a significant portion of the rest-frame FUV spectra is contaminated by features not related to stellar emission (e.g. interstellar absorption lines and nebular emission lines).
We excluded these wavelength regions, using only the spectral windows sensitive to photospheric absorption and stellar-wind features outlined in Table 3 of \citet{steidel2016}.
Finally, we also only considered wavelengths redward of \lya \ at $1216\rm{\AA}$ to avoid the additional uncertainties related to the intergalactic medium \hi \ absorption and dust correction at shorter wavelengths.
An example of a fit to one spectrum (VANDELS-m5) is shown in Fig. \ref{fig:example_fit}. 
We list the best-fitting log($Z_{\ast}$) values and errors in the last column of Table 1.
These quoted errors are the statistical uncertainties only, a discussion of the various systematic uncertainties, which we find to be at the $\simeq 10 \%$ level, is given in the appendix.

\subsection{Accuracy of the estimate and systematic uncertainties}\label{subsec:systematic_uncertainties}

Before discussing the stellar metallicities derived from the VANDELS data, it is worth considering how reliable we expect our method to be.
We performed two tests to explore any potential systematic uncertainties related to our adoption of a simplified star-formation and chemical-abundance history.
Firstly, we used data from the \fiby \ and \simba \ simulations to construct synthetic FUV spectra based on realistic star formation and chemical-abundance histories, and compared our recovered \logz \ values to the true FUV-weighted stellar metallicities.
Secondly, we derived stellar metallicities from the FUV spectra of local star-forming regions and starburst galaxies, which we compared to the published gas-phase metallicities of the same objects.
The results of both of these independent tests are discussed below.

    \begin{figure}
        \centerline{\includegraphics[width=\columnwidth]{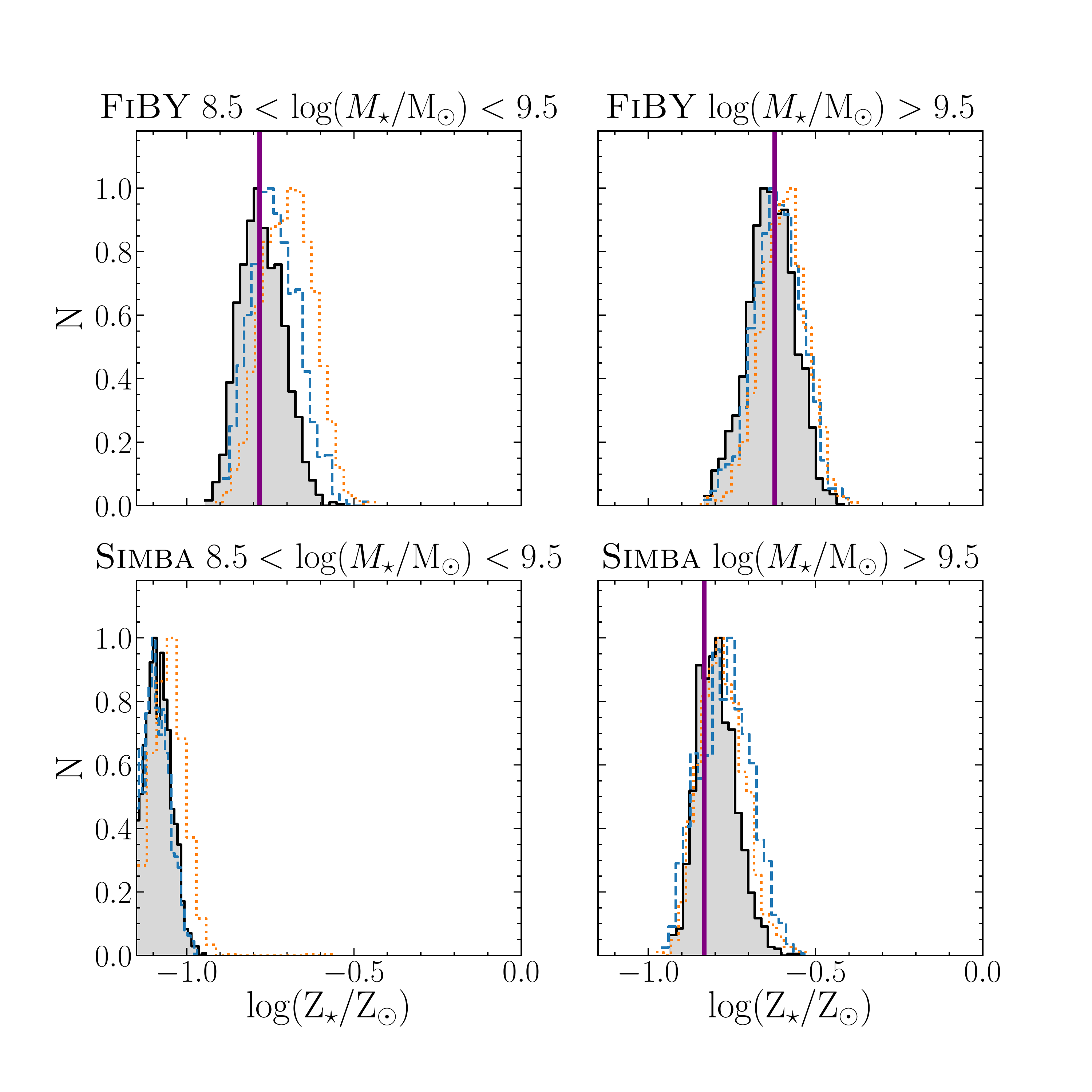}}
        \caption{The 1D posterior distribution for the UV luminosity weighted log($Z_{\ast}$/Z$_{\odot}$) recovered
         for four simulated galaxies.
        The simulated galaxies were built using the four average star-formation and chemical-abundance histories shown in Fig. \ref{fig:sim_sfr_chem} as described in the text.
        The simulation (\fiby/\simba) and stellar mass regime corresponding to each panel are given in the title.
        The grey filled histograms are the 1D posterior distributions obtained using our fiducial Starburst99 models assuming 100 Myr of constant star formation.
        The blue dashed and orange dotted curves are constant star formation models with longer timescales of 300 and 500 Myr.
        The purple vertical lines show the true UV-weighted metallicity of each simulated galaxy.
        For the lowest mass galaxies in \simba \ (lower left panel), the true UV-weighted metallicity is below the lower limit enforced by the Starburst99 models (log(Z$_{\ast}$/Z$_{\odot}$)$=-1.15$) and therefore not shown.
        In this case the fitted metallicity is, as expected, only an upper limit.}
        \label{fig:logz_recovery}
    \end{figure}

\subsubsection{Synthetic spectra based on simulation data}

In previous analyses of the FUV spectra of star-forming galaxies at high redshift, it has been common to assume models with constant star-formation histories \citep{rix2004,sommariva2012,steidel2016,strom2018}.
Generally, a star-formation timescale of ${100 \: \mathrm{Myr}}$ is assumed.
It is further assumed that the metallicity of the stars does not vary strongly across the star-formation history.
These assumptions make modeling the FUV spectra of high-redshift galaxies relatively simple compared to longer-wavelength regimes and/or lower-redshift galaxies, where the timescales involved are longer, so that more complex star formation and chemical evolution histories must be considered \citep[e.g.][]{carnall2018b,leja2018}.

Nevertheless, it is worth testing the validity of these assumptions against predictions from simulations.
To do this we have used the \fiby \ and \simba \ simulation data described in Section \ref{subsec:simulation_description}.
Because our composite spectra are averages across a population, we focused on the average star formation and chemical abundances histories of galaxies in the \fiby \ and \simba \ simulations. 
We considered two stellar mass bins which encompass the observed data, a low mass bin with $8.5 < $ \logm \ $< 9.5$ and a high mass bin with \logm \ $> 9.5$.
As a result of their different volumes the maximum mass is different in the two simulations ($10^{10.5}\rm{M}_{\odot}$ in \fiby \ and $10^{11.2}\rm{M}_{\odot}$ in \simba).
The average star formation and chemical evolution histories for these simulated galaxies are shown in Fig. \ref{fig:sim_sfr_chem}.

     \begin{figure*}
        \centerline{\includegraphics[width=7.5in]{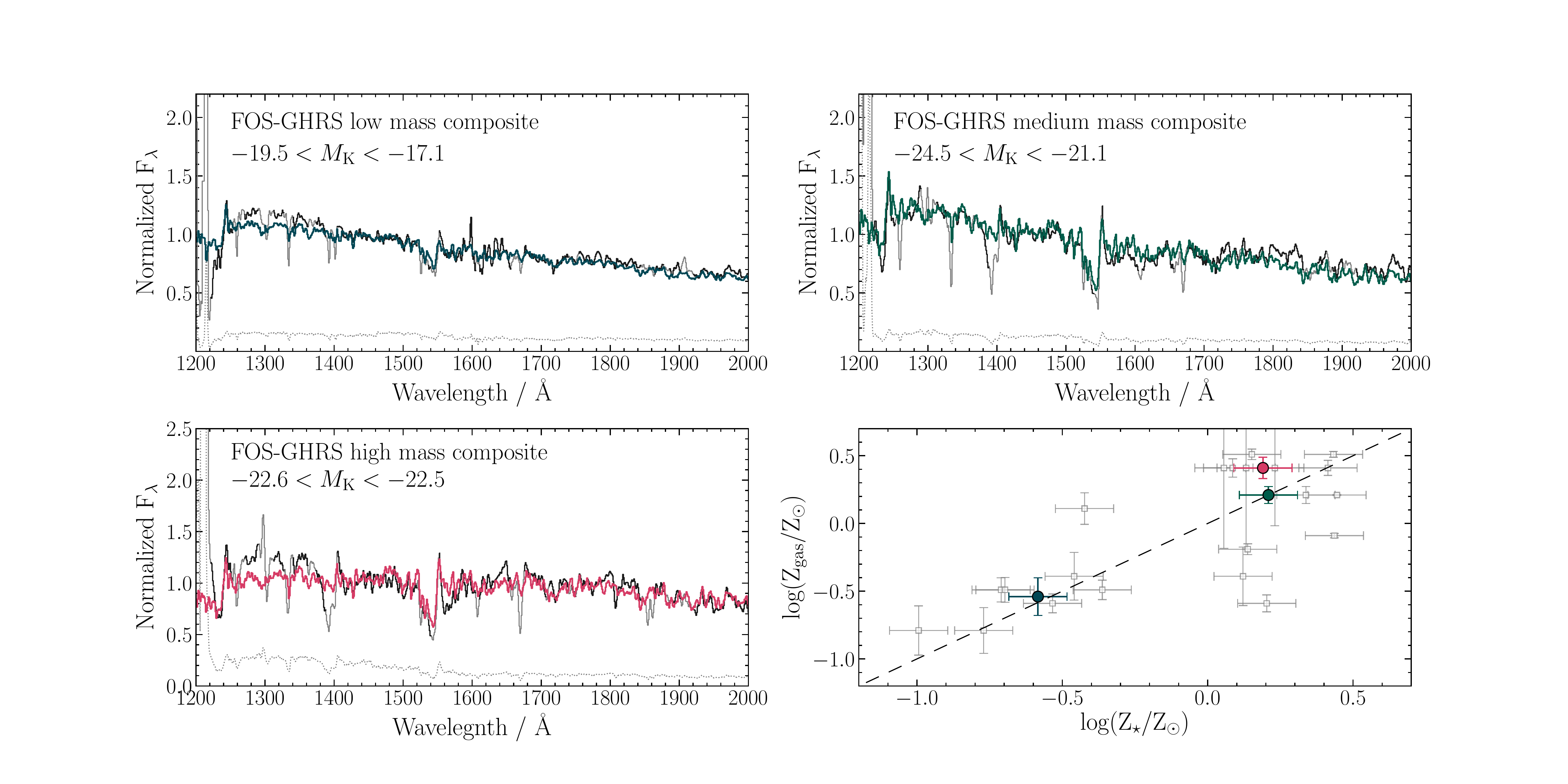}}
        \caption{Composite FOS-GHRS spectra in bins of absolute $K$-band magnitude ($M_{\rm{K}}$) with the best-fitting Starburst99 stellar population synthesis models over-plotted in colour (top two panels and lower left-hand panel).
        The lower right-hand panel shows the relationship between our derived stellar metallicities and published gas-phase metallicities for the FOS-GHRS galaxies, both for the individual galaxies (grey points) and the composites (coloured points), illustrating the expected correlation between these quantities (the Spearman rank correlation coefficient is 0.70).
        }
        \label{fig:fos_ghrs_fits}
    \end{figure*}

In accordance with common assumptions, it can be seen that, although the average star formation rates are smoothly rising over the full formation history, all are relatively constant within the past 100 \Myr.
Even in the case of the most extreme variation (the high-mass bin of \simba \ galaxies) the 100 \Myr \ averaged star formation rate is $29\pm4$ M$_{\odot}\rm{yr}^{-1}$ (ranging from $23 - 35$ M$_{\odot}\rm{yr}^{-1}$).
In fact, for the star formation rates, one could argue that there is relatively little variation over roughly $300-500$ \Myr \ timescales in all cases.
This is in good agreement with previous analyses of the average star formation histories of simulated galaxies at similar redshifts \citep{finlator2011}.

An invariance on timescales of $\simeq 100$ \Myr \ is also true for the chemical abundances, which overall follow a very similar trend to the star formation rates.
There is, in general, a slightly steeper rise over the longer timescales, although even in the most extreme case (again the high-mass bin of \simba \ galaxies) the increase is only a factor $\simeq 2$ over 500 \Myr.
Interestingly, despite the exact details of the global histories varying quite significantly, all share a common theme: rapid evolution over the first \Gyr \ (i.e. from formation up to $z\simeq6$) followed by a more gradual evolution over the next $500$ \Myr \ (i.e. $z=4-6$).
Based on the simulation data it would seem that assuming a simple constant star formation rate and chemical abundance model over $100$ \Myr \ timescales is a reasonable approximation for the composite galaxy spectra in our sample, since, across $100$ \Myr, \emph{both} the average star formation rates and chemical abundances of the galaxy population are relatively invariant in the two stellar mass bins.

However, below we will also demonstrate explicitly that our results should not be strongly affected by assuming longer timescales.
This is mainly due to the fact that, assuming constant star formation, the model FUV spectra reach an equilibrium (i.e. become time invariant) after $\simeq 50$ \Myr \ \citep[e.g.][]{leitherer2010}, and therefore any timescale $\gtrsim 50$ \Myr \ should yield similar results.
We also note that while these assumptions hold for galaxies on average, the same might not be true for individual galaxies.
On an individual basis, the star formation histories will not necessarily be smooth, and bursts of star formation will likely play a more significant role \citep[e.g.][]{hopkins2015}.

As an explicit test, we constructed four synthetic FUV spectra based on the four star-formation and chemical-abundance histories illustrated in Fig. \ref{fig:sim_sfr_chem}, using the SB99 WM-Basic instantaneous burst models scaled appropriately to match the integrated stellar mass.
To account for the effect of dust, we also attenuated each individual burst model according to its metallicity. 
Under our simple prescription, the absolute attenuation at $1500 \rm{\AA}$ is related to the metallicity via $A_{1500} = -2.24 \times \mathrm{log}(Z_{\ast}) - 2.16$, with a random scatter of $\sigma_{A_{1500}}=0.5$ magnitudes (in this case $Z_{\ast}$ is the total metallicity not the iron abundance).
This scaling relation is motivated by results from the FiBY simulation (Khochfar et al. in preparation), and implies $\simeq$ 2 magnitudes of FUV attenuation for stars formed within solar metallicity environments.
We note that, despite its simplicity, the relation yields global (i.e. galaxy averaged) $A_{1500}$ values in reasonable agreement with the the attenuation versus stellar mass relation presented in \citet{mclure2017}.
Finally, for simplicity, we used a Calzetti attenuation curve to convert $A_{1500}$ into $A_{\lambda}$ across the full FUV spectral range.
The synthetic spectra were smoothed to the resolution of the VANDELS data and gaussian noise was added assuming a S/N per pixel of 15 (comparable to the observed stacks).

One unfortunate aspect of this procedure is that, for a portion of all the star-formation histories, the metallicity is below the minimum imposed by the Starburst99 models ($Z_{\ast}/\mathrm{Z}_{\odot} = 0.07$; $\mathrm{log}(Z_{\ast}/\mathrm{Z}_{\odot}) = -1.15$).
In these cases we are forced to simply use the lowest metallicity SB99 spectrum.
This is particularly a problem for the low metallicities predicted in the low-mass bin of the \simba \ simulation (the black solid curve in the right-hand panel of Fig. \ref{fig:sim_sfr_chem}).
For the other three cases the issue is somewhat mitigated, since the FUV light is dominated by stars with metallicities within the SB99 parameter space.
Nevertheless, we note that this effect could still introduce some small bias into the recovered \logz \ values.

The results of the test are shown in Fig. \ref{fig:logz_recovery}, and generally confirm the conclusions drawn from Fig. \ref{fig:sim_sfr_chem}.
For the three cases in which the FUV-weighted metallicity is within the model parameter space the true value is well recovered, with derived (true) \logz \ values of $-0.78 \pm 0.06$ $(0.78)$, $-0.63 \pm 0.07$ $(0.62)$ and $-0.80 \pm 0.06$ $(0.83)$. 
As mentioned above, the low mass galaxies in the \simba \ simulation have an average FUV-weighted metallicity of ${\mathrm{log}(Z_{\ast}/\mathrm{Z}_{\odot})=-1.88}$, below the lower limit of the models.
In this case, obviously somewhat by design, the solution bumps up against the edge of the parameter space.
For a similar case within the observed sample we would consider this an upper limit on \logz.
It can also be seen that adopting constant star-formation rate models with longer timescales (300 and 500 \Myr) does not have a strong systematic affect on the recovered metallicity although, in general, the ${100 \: \mathrm{Myr}}$ models perform best.

Finally, we note that, although it is encouraging, this test cannot provide a definitive justification of our method, since we are building and fitting the FUV spectra with, fundamentally, the same set of SB99 models. 
However, it does serve to demonstrate that our simplified assumptions regarding the star formation and chemical-abundance histories should not significantly bias the recovered metallicity values.

\subsubsection{The FOS-GHRS local Universe sample}

    \begin{figure}
        \centerline{\includegraphics[width=\columnwidth]{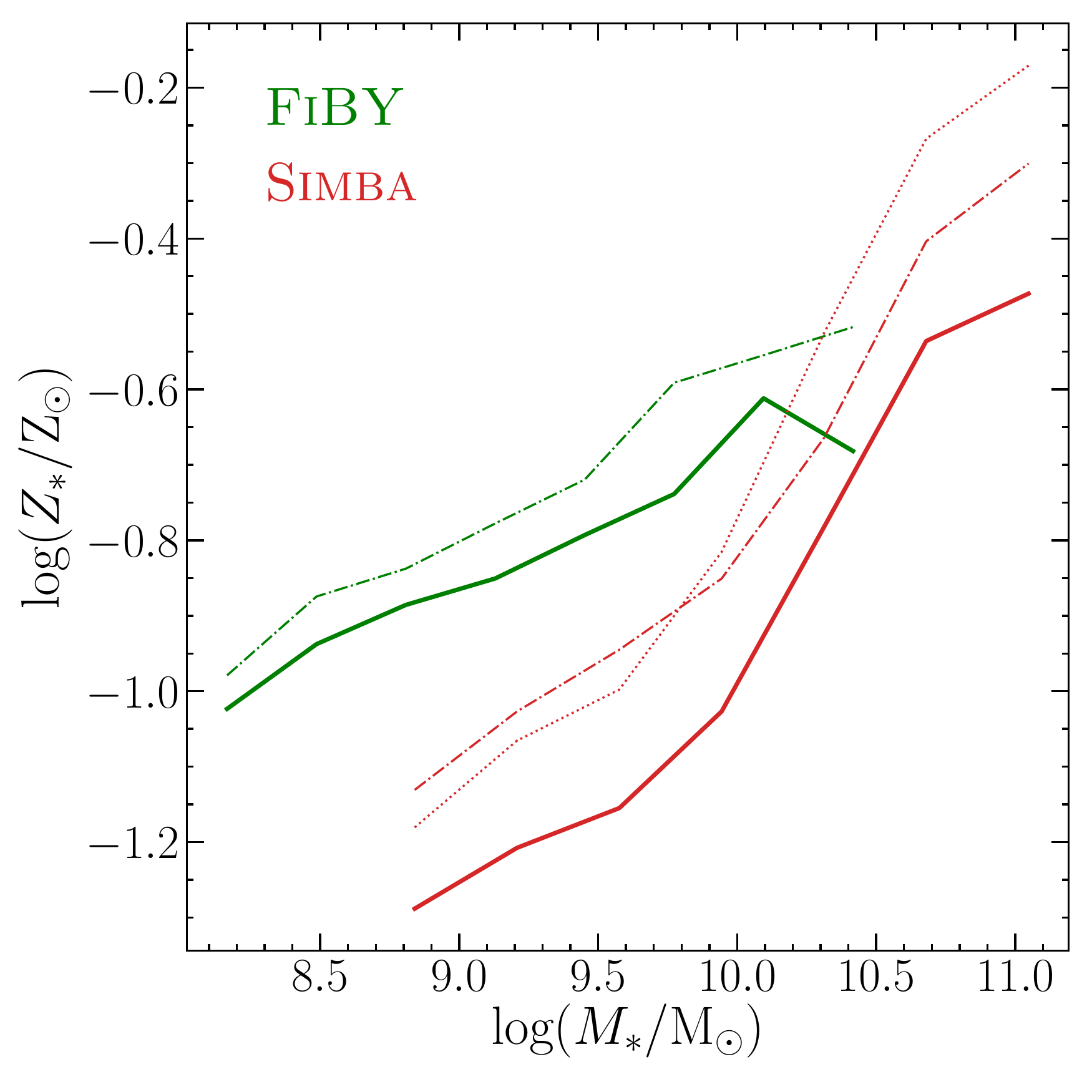}}
        \caption{The mass-weighted and FUV-weighted stellar mass-metallicity relations derived from the \fiby \ (green) and \simba \ (red) simulations at $z=4$.
        In each case the dot-dashed line shows the mass-weighted relation and the solid line shows the FUV-weighted relation, which is derived as described in the text (including a correction for dust).
        The FUV metallicities are generally offset to lower values by $\sim 0.05-0.1$ dex due to the preferential extinction of the youngest, most metal-enriched stars in the galaxy.
        For reference, the dotted-red line shows the intrinsic FUV-weighted relation (i.e. assuming no dust) for the \simba \ galaxies.}
        \label{fig:mw_vs_uvw_logz}
    \end{figure}

       \begin{figure*}
        \centerline{\includegraphics[width=7.5in]{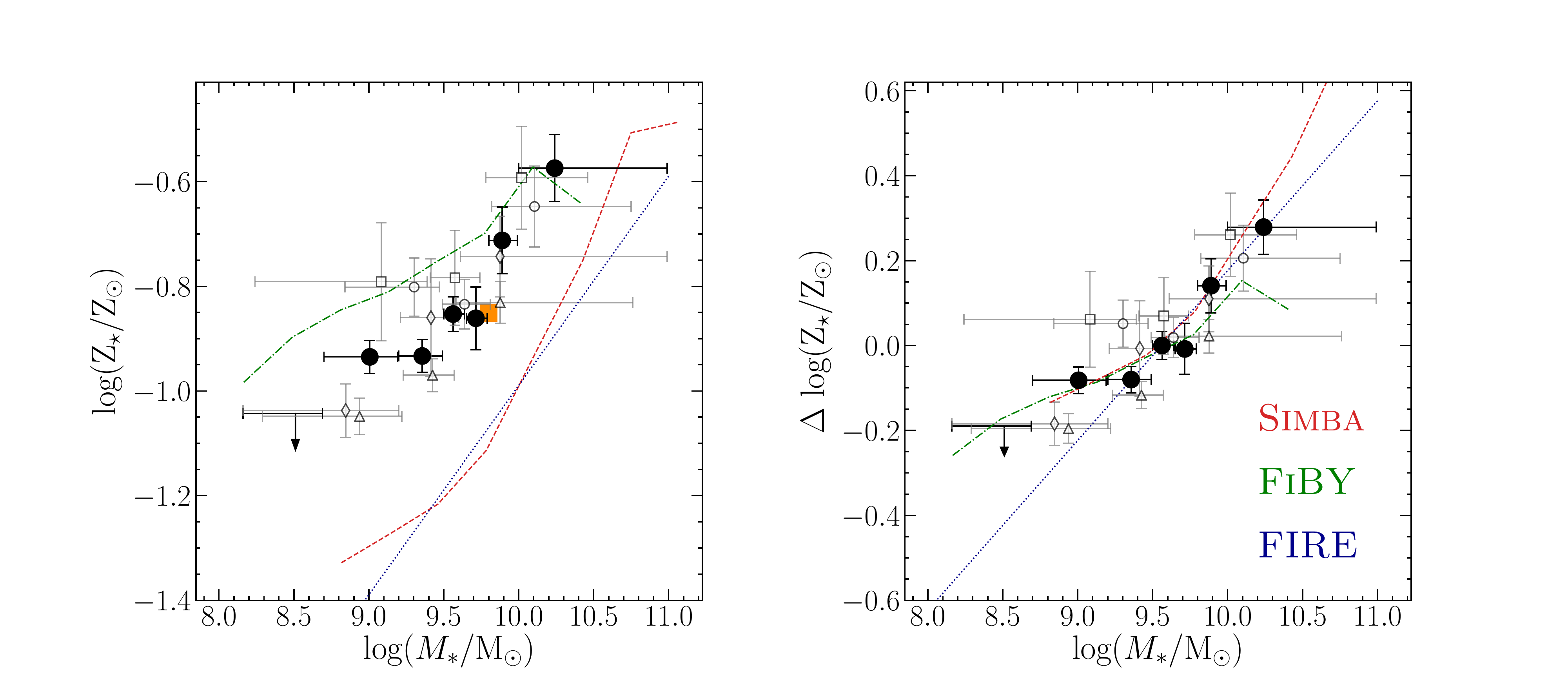}}
        \caption{The stellar mass-metallicity relationship for the VANDELS galaxies at $2.3 < z < 5.0$ compared to predictions from hydrodynamical simulations.
        The left-hand panel shows the absolute values of \logz \ plotted against \logm \ with the VANDELS data shown as filled black circles (the stellar mass-only composites) and open grey symbols (the composites binned in redshift and stellar mass).
        The redshift ranges corresponding to the open symbols are as follows: $2.3\leq z < 2.65$ (circles); $2.65\leq z < 3.15$ (squares); $3.15\leq z < 3.80$ (triangles); $3.80\leq z < 5.0c$ (diamonds).
        The vertical error bars show the 16th and 84th percentiles of the posterior distribution on \logz \ and the horizontal error bars show the range of stellar mass in each bin.
        The orange square data point is a measurement of \logz \ for a sample of 30 star-forming galaxies at $z=2.4$ taken from \citet{steidel2016}.
        The various lines are predictions for stellar metallicity (weighted by the $1500\rm{\AA}$ luminosity) versus stellar mass from three cosmological simulations: \fiby \ (green, dot-dashed), \simba \ (red, dashed) and FIRE (blue, dotted).
        The absolute values in all three simulations are inconsistent with the observed data.
        However, the shape of the relation is well recovered by the simulations as illustrated in the right-hand panel where, instead of absolute \logz, the values relative to \logm \ $\simeq 9.56$ are shown.}
        \label{fig:normalized_mzr}
    \end{figure*}

Given that the test described above, using synthetic spectra, is not truly independent, it is desirable to perform an independent test using real data.
To this end, we have used the above method to fit both the individual and composite spectra from the local FOS-GHRS galaxy sample described in Section \ref{subsec:fos_ghrs_data}.
The idea here is to compare the measured UV stellar metallicities to the published gas-phase metallicities of the same galaxies, under the assumption that the two parameters should correlate.
This should be a reasonable assumption, since the nebular emission lines used to determine the gas-phase metallicity are expected to originate from the hot, ionized, gas surrounding the same young, massive O- and B-type stars which dominate the FUV stellar emission.
Indeed, such a correlation is commonly observed in the \hii \ regions of Local Group galaxies, and is consistent with a 1:1 relation \citep[e.g.][]{toribio2017}.
Moreover, for the integrated emission of local galaxies (out to $z\simeq0.3$), a 1:1 correlation is observed for the young ($< 2$ \Gyr \ old) stellar population \citep[i.e. the population traced by rest-frame FUV observations;][]{gonzalezdelgado2014}.

As discussed in Section \ref{sec:data}, we produced three composite FOS-GHRS spectra in bins of absolute K-band magnitude ($M_{\rm{K}}$, a proxy for the stellar mass) as outlined in Table \ref{table:fos_ghrs_sample_appdx}.
Spectral fits to the $M_{\rm{K}}$ composite spectra are shown in Fig. \ref{fig:fos_ghrs_fits}.
We find that the stellar metallicities of the two brightest composites ($M_{\rm{K}}= -23.1$ and $-25.1$; or ${\mathrm{log}(M_{\ast}/\mathrm{M}_{\odot})=10.33}$ and $11.13$) are similar with \logz \ $\simeq 0.2$ (or $Z/Z_{\odot}\simeq1.6$) in excellent agreement with their median gas-phase metallicities.
For the faint, low-mass, composite with $M_{\rm{K}}= -18.5$ (\logm \ $=8.5$) we find a stellar metallicity of \logz \ $\simeq -0.58$ (or $Z/Z_{\odot}\simeq0.26$), roughly a factor six lower) and, again, the stellar metallicity is in good agreement with the median gas-phase metallicity.
We note that the fits to the low-mass and high-mass composites, while generally good, do not appear to match the observed spectra in the range $\simeq 1200-1400 \rm{\AA}$ despite having statistically acceptable reduced $\chi$-squared values ($\chi^2_r \simeq 0.7$).
These fits would likely be improved with a more complex dust model, as the current discrepancy could plausibly be a result of strong variation in the dust law at shorter wavelengths due to the relatively low number of objects in each composite. 
However, as the current fits are statistically acceptable, we did not consider a more complex dust model in this analysis.

We find a similar result for the individual galaxies, which are fitted in the same way (although often using only a restricted portion of the full wavelength range), albeit with a much larger scatter.
In fact, the results in Fig. \ref{fig:fos_ghrs_fits} are consistent with the stellar and gas-phase metallicities scattering about a 1:1 relation.
The Spearman rank correlation coefficient between the two quantities is $0.70$, with the probability of obtaining this value by chance formally zero ($p=3.1 \times 10^{-4}$).
We note that although a correlation between stellar and gas-phase abundances is to be expected, a direct 1:1 correspondence is only expected if the O/Fe ratios (i.e. chemical abundance patterns) in these galaxies are close to the solar value, we will return to this issue in the case of high-redshift galaxies in Section \ref{subsec:o_fe_ratios}.

In summary, we have performed two independent tests of our adopted method.
Firstly, we have shown that the assumption of a constant star-formation history with a timescale of 100 \Myr, at a single metallicity, is sufficient to recover the FUV-weighted stellar metallicities of synthetic galaxy spectra built from more realistic star formation and chemical-abundance histories.
Furthermore, by applying our method to composite FUV spectra of local galaxies we have shown that we are capable of reproducing the expected trend between stellar and gas-phase abundances.
One further potential source of bias which we do not discuss in detail here, relates to the effect of redshift uncertainties on the metallicity-sensitive continuum features when constructing composite spectra. 
However, we do not expect this effect to significantly affect our derived metallicities, and a brief discussion is provided in Appendix \ref{sec:app_redshift_uncertainties}.

 	\begin{figure*}
        \centerline{\includegraphics[width=6.0in]{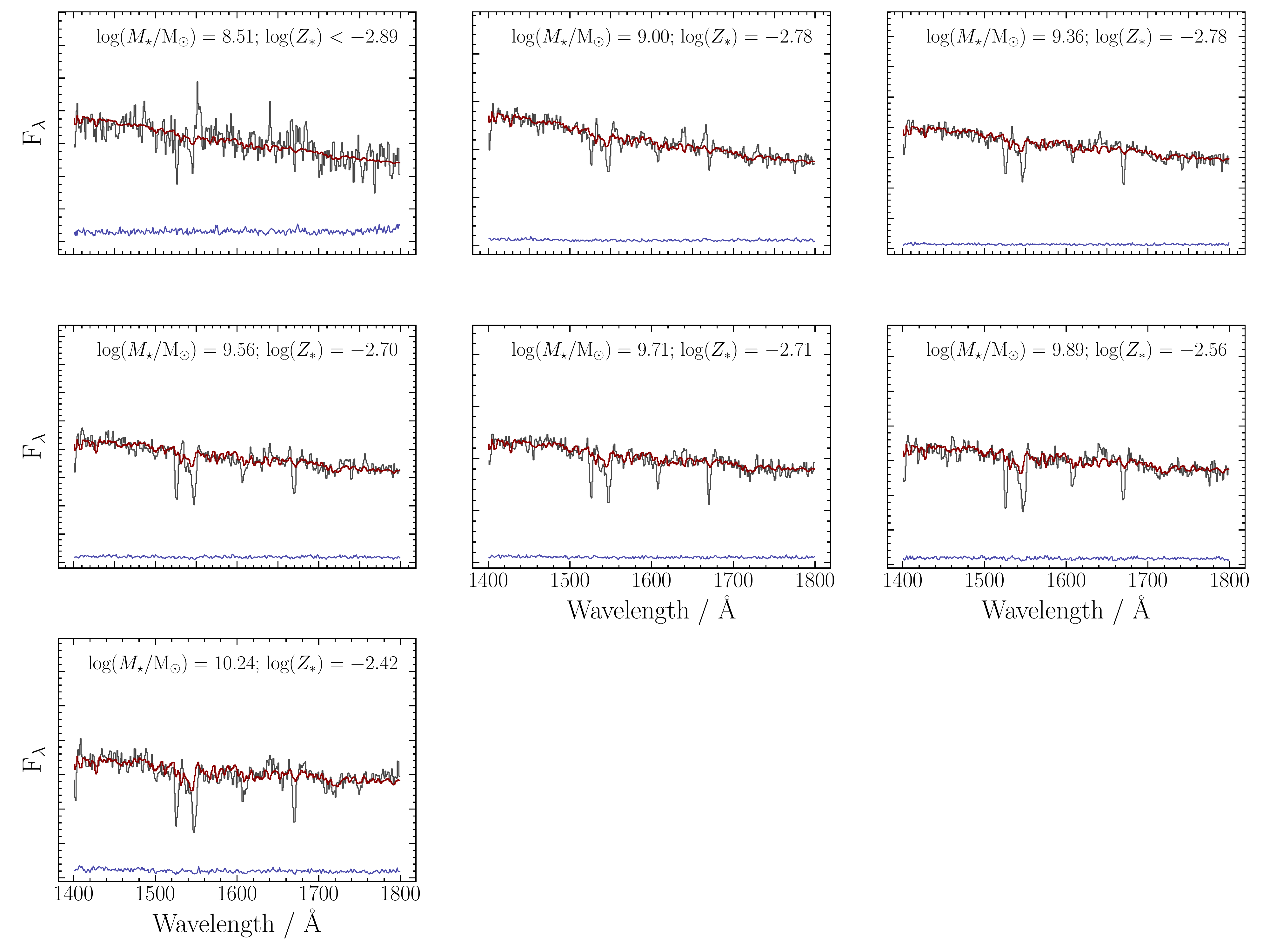}}
        \caption{Fits to the VANDELS stellar-mass composite spectra at $2.5 < z < 5.0$. 
        Each panel shows one of the seven composites with the average stellar mass and best-fitting metallicity indicated in the legend. 
        The black curves show the observed spectra with the error spectra shown in blue. 
        The best-fitting Starburst99 WM-Basic spectra are over-plotted in red.
        The figure focuses on the wavelength region $1400 < \lambda < 1800$ to illustrate how the increasing metallicity at higher stellar masses causes more pronounced undulations in the continuum.
        In addition to this, the equivalent widths of metallicity-dependent stellar absorption features such as \niv \ and \civ \ can be seen to decrease at lower stellar masses.}
        \label{fig:mass_stack_fits}
    \end{figure*}

\section{The Stellar Metallicities of Star-Forming Galaxies at $\mathbf{2.5 < z < 5.0}$}\label{sec:results_stellar_mzr}

In this section we discuss stellar mass-metallicity relationship at $2.5 < z < 5.0$ derived from the VANDELS composite spectra.
Before discussing the observational results however, it is important to consider that fact that the stellar metallicities derived using our method are FUV-weighted, and it is therefore worth briefly discussing how FUV-based metallicities (tracing the recent star-formation history) might be biased with respect to the mass-weighted metallicities (which trace the global star-formation history) at these redshifts.  

    \begin{figure*}
        \centerline{\includegraphics[width=5.5in]{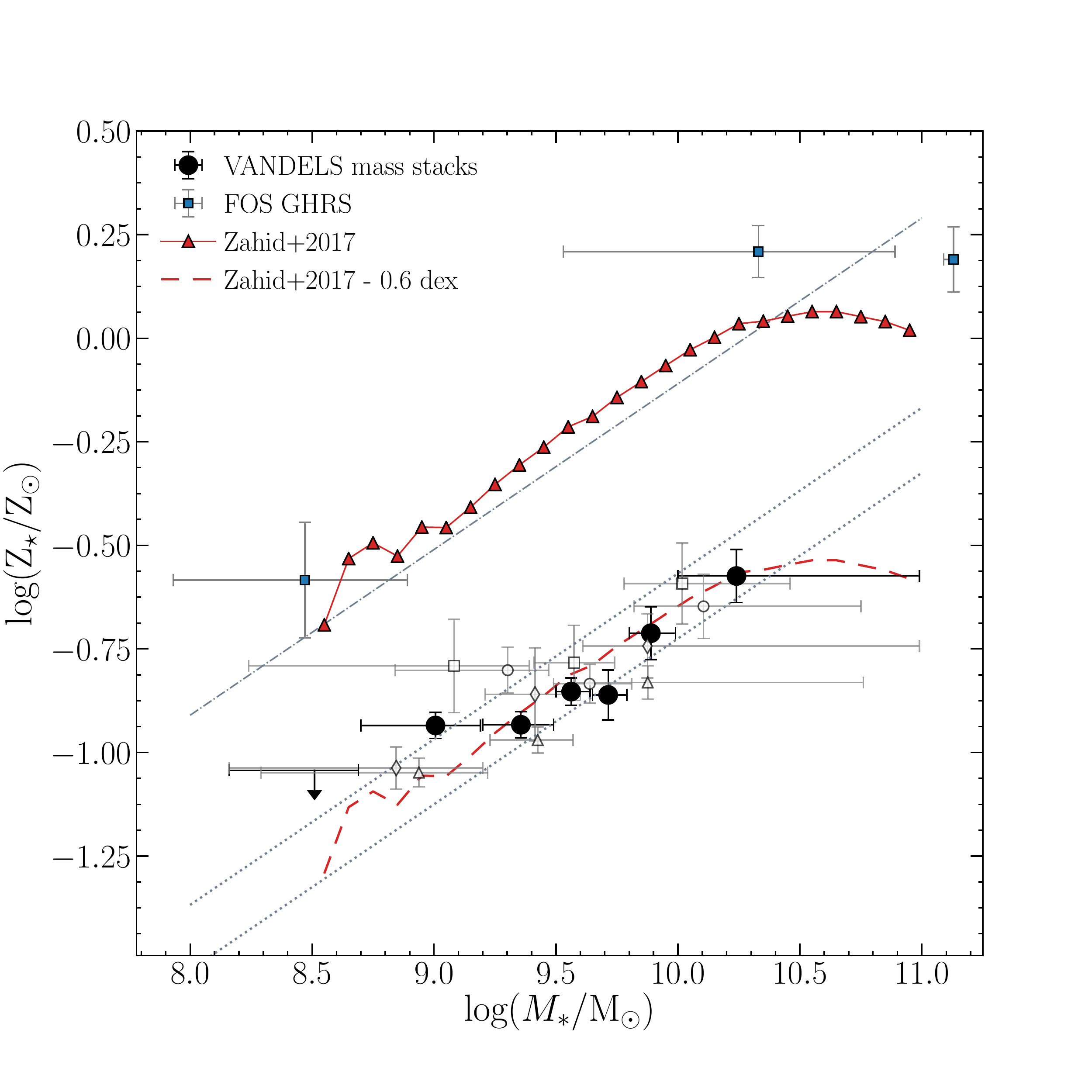}}
        \caption{Redshift evolution of the stellar mass versus stellar metallicity relationship.
        The red triangular data points show the $z=0$ relation derived from fitting stacked optical continuum spectra of $\sim 200,000$ star-forming galaxies in the Sloan Digital Sky Survey \citep{zahid2017}.
        The blue square data points show the FUV-based stellar metallicities for the composite FOS-GHRS spectra of local star-forming regions and starburst galaxies discussed in the text.
        The VANDELS data at $2.5 < z < 5.0$ are shown as the black/grey data points as in Fig. \ref{fig:normalized_mzr}.
        The dashed red line shows the $z=0$ relation of \citet{zahid2017} shifted by $-0.6$ dex in \logz.
        Simply shifting the local relation down by a factor $\simeq 4$ produces remarkably good agreement with the high redshift data.
        The dotted grey lines show the FIRE prediction at $z=2.3$ (upper) and $z=5.0$ (lower) and the grey dot-dashed line is the FIRE prediction at $z=0$.
        The FIRE metallicities have been arbitrarily scaled upwards by a factor of 2 (0.3 dex).}
        \label{fig:stellar_mzr_allz}
    \end{figure*}

\subsection{FUV weighted versus stellar-mass weighted metallicities}

From the \fiby \ and \simba \ simulation data, we were able to derive both the mass-weighted and FUV-weighted stellar metallicities for each simulated galaxy and we present the results of the comparison in Fig. \ref{fig:mw_vs_uvw_logz}, which shows the median mass-weighted and FUV-weighted stellar mass-metallicity relations for both simulations.
The mass-weighted metallicities of each simulated galaxy were derived simply using the mass and metallicity information of each star particle.
For the FUV-weighted metallicities we used the FUV luminosity of each star particle at $1500 \rm{\AA}$ taken from the Starburst99 instantaneous burst models.
In both cases, the FUV-weighted relation is offset to lower metallicity, although crucially the shape of the relation is unaffected.
The offset is due to the fact that the youngest stars, which dominate the FUV flux, are generally more metal-enriched and therefore, according to our attenuation prescription, are more heavily attenuated and do not contribute as much to the FUV.
For example, it can be seen from Fig. \ref{fig:mw_vs_uvw_logz} that the intrinsic FUV-weighted relation (i.e. assuming no dust attenuation) for the \simba \ galaxies  (red-dotted line), closely follows the mass-weighted relation.
Therefore, any preferential attenuation of the youngest, most metal-enriched, stars will result in a bias to lower metallicities when using FUV spectra.
There is clearly some systematic uncertainty here related to our method of attenuating the individual star-particle spectra, however, based on reasonable dust assumptions, we expect the FUV-weighted metallicities of high-redshift star-forming galaxies to be biased low with respect to the mass-weighted values by $\simeq 0.05-0.1$ dex, but predict that the shape of the mass-metallicity relation to be similar in either case.
Finally, we note that throughout this section we will also compare our results to the relations extracted from the FIRE simulation presented in \citet{ma2016}\footnote{\citet{ma2016} assume a solar metallicity of 0.02 and therefore we have converted their relations to our assumed value of 0.0142.}.
Since these FIRE relations are for mass-weighted metallicities, we have scaled them down by 0.05 dex.

\subsection{The $\mathbf{2.5 < z < 5.0}$ stellar mass-metallicity relation}

The $2.5 < z < 5.0$ stellar mass-metallicity relation derived from the VANDELS sample is shown in Fig. \ref{fig:normalized_mzr}.
The vertical error bars show the 16th and 84th percentiles of the posterior distribution on \logz \ and the horizontal error bars show the range of stellar mass in each bin.
The first thing to note is that a clear monotonic increase in \logz \ with \logm \ is observed for our sample.
Focusing on the purely stellar-mass stacks, \logz \ evolves from a lower limit of $< -1.04$ $(68\%)$ at \logm \ $\simeq 8.5$ to $-0.57 \pm 0.06$ at \logm \ $\simeq 10.2$, an increase by a factor $\geq3$ in the average metallicity across roughly two decades of stellar mass. 
In Fig. \ref{fig:mass_stack_fits} we show how this metallicity evolution affects the form of the FUV spectrum by showing the best-fitting SB99 models overlaid on the composite spectra (the effect is subtle, but as the average mass of the composite spectra increases, the continuum becomes less smooth and the undulations resulting from metal line blanketing become more pronounced).
At all masses, the stellar metallicities (to first order a proxy for the iron abundance) are significantly sub-solar $(\lesssim 25\%)$. 
From an analysis of the redshift and mass tacks, we do not find strong evidence for a significant trend with redshift at a given stellar mass in the VANDELS sample. 
It can be seen from Fig. \ref{fig:normalized_mzr} that the redshift-mass stacks are consistent with scattering around the stellar mass-only relation.

With respect to the normalization of the stellar mass-metallicity relation, our results are consistent with recent results from the independent KBSS-MOSFIRE survey \citep{steidel2016,strom2018}.
The orange data point in Fig. \ref{fig:normalized_mzr} shows the result derived from a stack of 30 star-forming galaxies at $z=2.40\pm0.11$ presented in \citet{steidel2016} who estimated $Z_{\ast}/\mathrm{Z}_{\odot} \simeq 0.14$ using a similar method to our own, by fitting Starburst99 WM-Basic models to the composite FUV spectrum, with the additional requirement that the best-fitting stellar model should also account for the observed nebular optical emission lines (a constraint we currently do not have for our sample).
\citet{steidel2016} find a lower best-fitting metallicity model using the alternative BPASSv2 stellar models, which is consistent with what we report in Appendix \ref{sec:bpassv2_comp}. 
Despite the slight difference in method, average redshift, and the stellar models employed, the consistency is encouraging, and provides further evidence that the stellar metallicities (or iron abundances) of galaxies at high redshift are significantly sub-solar ($Z_{\ast}/\mathrm{Z}_{\odot} \lesssim 0.25$).

\subsection{Redshift evolution of the stellar mass-metallicity relation}

Extending the comparison to lower redshifts, we show the evolution of the mass-metallicity relation between $2.3 < z < 5.0$ $(\langle z \rangle = 3.5)$ and $z=0$ in Fig. \ref{fig:stellar_mzr_allz}.
For a our main local Universe comparison sample we have used a recent determination of the stellar mass-metallicity relation for $\sim 200,000$ star-forming galaxies in the Sloan Digital Sky survey by \citet{zahid2017} (Z17).
The solid line (connecting the triangular data points) in Fig. \ref{fig:stellar_mzr_allz} shows the $z=0$ relation and the dashed line shows an arbitrary downward shift of this relation by 0.6 dex.
This simple shift matches the VANDELS data remarkably well, implying an increase in metallicity of a factor $\sim4$ from $z \sim 3.5$ to the present day without any obvious dependence on stellar mass.

However, it is important to note that there will be some biases associated with this simple comparison. 
For example, whereas our metallicities are determined from the rest-frame FUV, tracing the massive OB stellar population, and are sensitive to the iron abundance, \citetalias{zahid2017} derive metallicities by fitting to rest-frame optical spectra in the wavelength range $\simeq 4000 - 7000 \rm{\AA}$.
It is not immediately obvious that the methods trace either the same stellar population or abundance type.
We can partially address this issue via a comparison with the metallicities derived from the composite FOS-GHRS spectra (blue data points in Fig. \ref{fig:stellar_mzr_allz}).
In this case we know that the method, at least, is not biasing the comparison since the metallicities have been derived in identical ways.
Although the mass sampling is not ideal, from the three stellar-mass bins available we find the derived stellar metallicities are in approximate agreement with the $z=0$ relation, especially when taking into account that, when measured for individual galaxies, the scatter in the local stellar mass-metallicity relation is of the order $\gtrsim 0.1$ dex \citep{panter2008}.
There are again a number of caveats, primarily the fact that, for the high-mass galaxies, the FOS-GHRS observations are often taken for individual \hii \ regions within the galaxy rather than the galaxy globally.
However, one can make the argument that these regions are likely to dominate the global FUV emission of these galaxies.
Taken as a whole, the data presented in Fig. \ref{fig:stellar_mzr_allz} provide strong evidence for an evolution in stellar metallicities of roughly a factor 4 between $2.3 < z < 5.0$ and $z=0$.

Another potential complication worth acknowledging is the fact that, although comparing the stellar metallicities of star-forming galaxies at these redshifts is clearly of interest, it is likely that the two datasets do not form an evolutionary sequence since, depending on their stellar mass, star-forming galaxies at high redshift will likely be the progenitors of present-day quiescent galaxies \citep[e.g.][]{carnall2018}.
Nevertheless, the stellar metallicities of early-type galaxies with $M_{\ast}>10^{10}\rm{M}_{\odot}$ estimated from rest-frame optical absorption features also fall predominantly within the range $-0.2<$ \logz \ $<0.2$ \citep{gallazzi2006}, similar to the values for the star-forming population, and therefore a similar conclusion would be drawn from this comparison.

In summary, we find clear evidence for a monotonic increase in stellar metallicity with galaxy stellar mass across the redshift range $2.3 < z < 5.0$.
Within this narrow redshift window we find no strong evidence for redshift evolution at fixed stellar mass.
However, a direct comparison to the stellar metallicities of star-forming galaxies in the local Universe implies that metallicities increase by a factor of 4, independent of mass, between $z \sim 3.5$ and $z=0$.

\section{Discussion}\label{sec:discussion}

Here we discuss some of the implications of our results.
Firstly, we compare the observed data to the predictions of three cosmological simulations.
Then, using a simple one-zone chemical evolution model, we attempt to identify the key physical parameters driving both the significantly sub-solar stellar metallicities, as well as the observed mass dependence in our sample.
Finally, we compare the stellar metallicities derived here with published gas-phase metallicities at similar redshifts in order to investigate suggestions of enhanced $\mathrm{O}/$Fe ratios in star-forming galaxies at $z \gtrsim 2.5$.

\subsection{Comparison to cosmological simulations}\label{subsec:sim_discussion}

The left-hand panel of Fig. \ref{fig:normalized_mzr} compares the absolute \logz \ values to simulation predictions for the FUV-weighted stellar mass-metallicity relation. 
In general, the absolute metallicity values predicted in the simulations are either systematically too large (\fiby) or too small (\simba \ and FIRE).
Given these discrepancies are generally a factor $\lesssim 2$, they could be explained by uncertainties in the element yields from core-collapse supernovae (CCSNe), SNe Ia and asymptotic giant branch (AGB) star winds used in the simulations \citep[e.g.][]{romano2010}.
Indeed, yields in the cosmological simulation \textsc{Mufasa} \citep[][the precursor to the \simba \ simulations]{dave2016} were arbitrarily reduced by a factor of two to match published gas-phase oxygen abundance data.
The correction became unnecessary in \simba \ due to a more detailed model for how elements become locked in dust grains \citep{dave2019}, however this comparison would suggest that the iron-peak element yields may still not consistent with current data.
Another explanation, illustrated in Section \ref{subsec:waf} below, could be related to the strength of galactic outflows in the simulations.

The shape of the relation, on the other hand, appears to be relatively well reproduced by all three simulations.
The right-hand panel of Fig. \ref{fig:normalized_mzr} shows the data and simulated relations normalized at \logm \ $= 9.56$ (the stellar mass of the VANDELS-m4 composite).
The exact slope of the relation based on the observed data is somewhat uncertain given the upper limit at the lowest stellar mass bin, but the simulations are clearly in relatively good agreement.
The \simba \ data, in particular, match the shape of the relation extremely well.

Determining what governs the overall shape of the stellar mass-metallicity relation in simulations is not straightforward due to the complex interplay of star formation, inflows and outflows which determine the metallicity evolution.
However, as we discuss in Section \ref{subsec:waf} below, one of the key physical parameters governing the shape and normalization of the relation is the strength and mass-scaling of galactic outflows, parameterized by the mass-loading parameter $\eta = \dot{M}_{\rm{outflow}}/\dot{M}_{\ast}$. 
The prescription for how the mass-loading parameter scales with stellar mass (the $\eta-M_{\ast}$ relation) in \simba \ is, in fact, based on the results of the high-resolution FIRE zoom simulations, which include a detailed stellar feedback model accounting for the effects of radiation pressure, supernovae, stellar winds and photoionization and photoelectric heating \citep{hopkins2014,muratov2015,angles-alcazer2017}.
The fact that \simba \ and FIRE use a similar outflow model could account for the good agreement in the shape of the MZR between the two simulations.
The feedback model used in \fiby \ is based on an earlier prescription for supernova feedback \citep{dalla_vecchia2012} which results in generally weaker stellar winds (i.e. smaller mass-loading parameters), although crucially the scaling of $\eta$ with stellar mass is similar to the \simba \ and FIRE simulations (S. Khochfar private communication).
The similarity of the slope of the $\eta-M_{\ast}$ relation in each simulation could therefore account for the consistency in their predictions for the shape of the MZR. 
On the other hand, the differences in the absolute \logz \ values are likely to be a result of both the different normalizations of the $\eta-M_{\ast}$ relations, as well as differences in the assumed stellar yields.

Finally, we note that the lack of observed redshift evolution between $2.5 < z < 5.0$ is also in reasonable agreement with predictions from simulations. 
For example, the FIRE simulation predicts $\simeq 0.16$ dex of evolution in \logz \ at fixed \logm \ across $2.3 < z < 5.0$ \cite[${1.7 \: \mathrm{Gyr}}$ of cosmic time;][]{ma2016}, a value roughly comparable to the $1-2\sigma$ error bars on the derived metallicities for the mass-redshift composites.
This is shown by the grey dotted lines in Fig. \ref{fig:stellar_mzr_allz}, where we have scaled the FIRE metallicities upwards by a factor of 2 such that the $z=2.3$ and $z=5.0$ relations bracket our data.
However, given the fact that we are only utilizing $\sim 65 \%$ of the full VANDELS dataset, evidence for a redshift evolution will be an area that can be explored in more detail in future work. 
Fig. \ref{fig:stellar_mzr_allz} also shows the prediction for the redshift evolution of the stellar mass-metallicity relation to $z=0$ from the FIRE simulation\footnote{Unfortunately, we cannot produce similar predictions from the \fiby \ and \simba \ simulations since they terminate at redshifts $z=4$ and $z=1$ respectively.}.
Again, it can be seen that the predicted evolution of the stellar MZR by $\simeq 0.6$ dex is in remarkable agreement with the observations.
We also note that an evolution of this magnitude is predicted from the EAGLE simulations \citep[][]{derossi2017}, although we do not show this relation in Fig. \ref{fig:stellar_mzr_allz} for clarity.
In summary, the current predictions from cosmological simulations seem to be in broad agreement with our data.

\subsection{Analytic one-zone chemical evolution models}\label{subsec:waf}

Although simple analytic models lack the rigor of detailed cosmological simulations, they can still provide useful physical insights into galactic chemical evolution. 
The primary advantage of using analytic approximations is, perhaps, the ability to quickly determine the sensitivity of chemical evolution to different physical parameters, a procedure which is computationally expensive in the case of detailed simulations.
Such analytic models are typically specified by a gas-accretion history, a star-formation law, and prescriptions for gas inflows/outflows and nucleosynthetic yields.
Numerous analytic models exists \cite[e.g.][]{lilly2013,zhu2017} however,
in this section, we compare our results with the analytic one-zone chemical evolution model presented in \citet{weinberg2017} (hereafter \citetalias{weinberg2017}).
The \citetalias{weinberg2017} model is particularly suited to our purpose since, as discussed, the FUV stellar metallicities we derive are a proxy for the iron abundance, and \citetalias{weinberg2017} incorporate a realistic delay time distribution (DTD) for Type$-$Ia SNe, allowing the evolution of iron peak elements to be tracked separately to that of the $\alpha$ elements.

    \begin{figure*}
        \centerline{\includegraphics[width=7.0in]{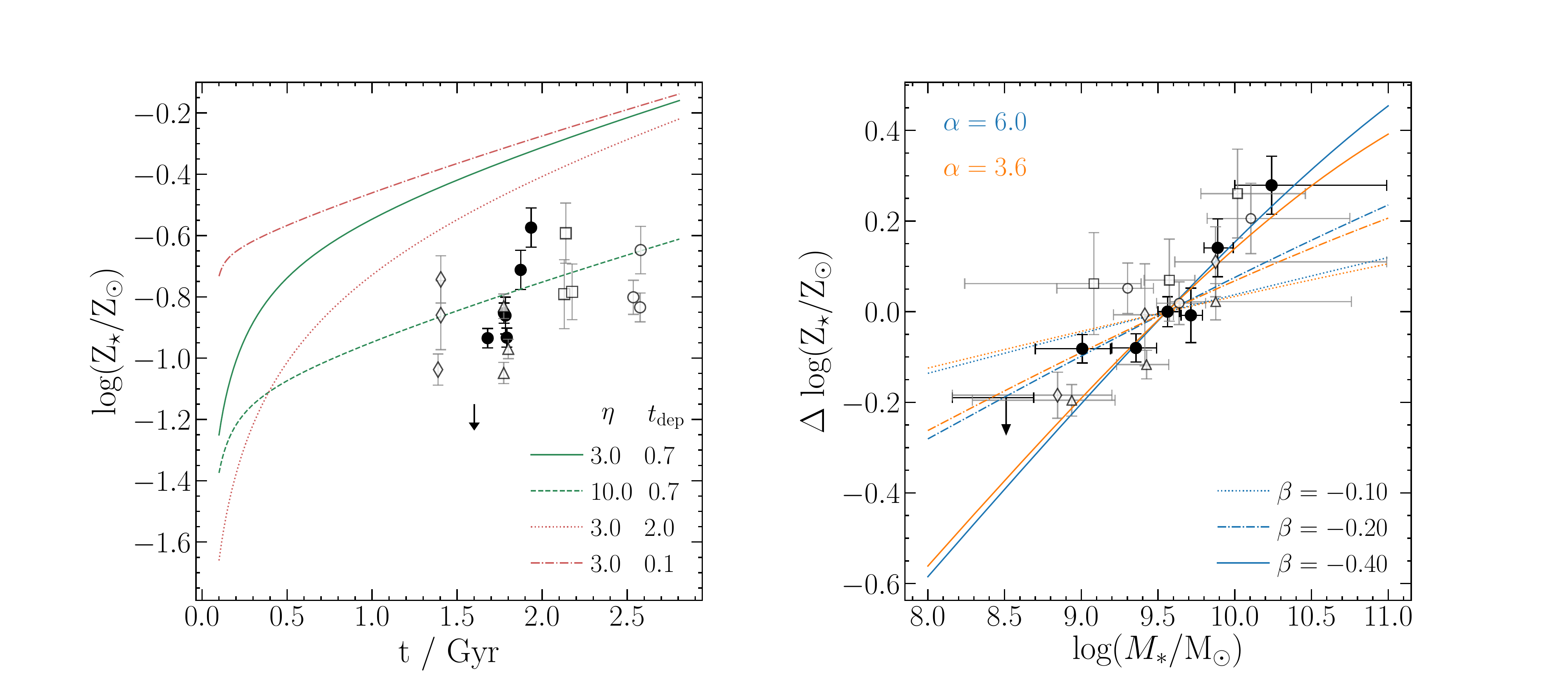}}
        \caption{The evolution of \logz \ as a function of cosmological time (left-hand plot) and $\Delta$\logz \ as a function of stellar mass (right-hand plot) compared to predictions from the one-zone analytic chemical evolution models of \citet{weinberg2017}. 
        The left-hand plot shows the time evolution of \logz \ from $t=0$ (i.e. the Big Bang) to the age of the Universe at the minimum redshift of our sample ($z=2.3; t=2.8$ Gyr).
        The green curves show \logz \ evolution assuming a constant depletion timescale ($t_{\rm{dep}}=0.7$ \Gyr) for two different values of the mass-loading parameter ($\eta=3$ and $10$) while the maroon curves show the evolution assuming a fixed value of $\eta=3.0$ and two different values of the depletion timescale ($t_{\rm{dep}}=0.1$ and $2$ \Gyr).
        The significantly sub-solar abundances of our sample ($\sim 10-20\%$) require mass outflow rates roughly an order of magnitude greater than the star formation rates (averaged across all stellar masses).
        The right-hand plot shows the difference in \logz \ as a function of stellar mass assuming different forms of the $M_{\ast}-\eta$ relationship.
        The functional form is taken from the scaling relation presented by \citet{muratov2015} (see Eq. \ref{eq:mass_loading_param_scaling}).
        The orange and blue lines show the $\Delta$\logz-\logm \ predictions relative to the value at \logm$\simeq9.5$ for different assumptions of the absolute mass-loading parameter at that stellar mass; $\eta=3.6$ and $\eta=6$ (see text for discussion).
        Regardless of the normalization, the evolution of \logz \ as a function of stellar mass is consistent with a power law exponent $\beta <-0.2$ and in good agreement with the \citet{muratov2015} result ($\beta=-0.36$).
        }
        \label{fig:waf17_models}
    \end{figure*}

Briefly, the key elements of the \citetalias{weinberg2017} model are as follows.
The star-formation rate ($\dot{M}_{\ast}$) is assumed to be proportional to the gas mass ($M_{\rm{gas}}$) with the star formation efficiency (SFE $=\dot{M}_{\ast}/M_{\rm{gas}}$) assumed to be constant (the so-called `linear Schmidt law').
For the purposes of this discussion we will refer to the inverse of the star-formation efficiency as the gas depletion timescale ($t_{\rm{dep}}=M_{\rm{gas}}/\dot{M}_{\ast}$) as this is a quantity that has been extensively studied out to $z\simeq6$ \citep[e.g.][]{scoville2016}.
Gas outflows are assumed to be a constant multiple of the star-formation rate, with the constant of proportionality commonly referred to as the mass-loading parameter ($\eta = \dot{M}_{\rm{outflow}}/\dot{M}_{\ast}$).
Therefore, for a specified star-formation history, the gas depletion timescale and mass-loading parameter define the total gas-accretion history.
The gas accreted from the IGM is assumed to be pristine ($Z=0$) and the outflowing gas is assumed to be at the same metallicity as the ambient star-forming ISM\footnote{We note that this is one important simplifying assumption of the model, since there is strong evidence that outflows are in fact metal-enriched with respect to the average ISM metallicity \citep[e.g.][]{peeples2011,chisholm2018}.}.
Three star-formation histories are explicitly solved for in \citetalias{weinberg2017} but we only consider the linear$-$exponential case here\footnote{In the linear-exponential case, the star-formation history is modeled as the product of a linear rise and exponential decline: $\dot{M}_{\ast}(t) \propto te^{\sfrac{-t}{\tau_{\rm sfh}}}$.}.
For our purposes, this essentially equates to a linearly rising star-formation history.
Elements produced by core-collapse supernova (CCSNe) are instantaneously recycled into the star-forming ISM, whereas enrichment from Type Ia SNe is delayed as specified by the DTD.
The form of the DTD is exponentially decreasing in time with a minimum delay time of 0.15 Gyr and an $e$-folding timescale of 1.5 Gyr\footnote{Again, it should be noted that this DTD is not necessarily consistent with current observational data \citep[e.g.][]{maoz2014} and represents another source of systematic uncertainty when considering the predictions of the \citetalias{weinberg2017} model.}.
The models assumes a \citet{kroupa2001} IMF and the CCSNe yields of \citet{chieffi2004} and \citet{limongi2006} which predict that $1.5\mathrm{M}_{\odot}$ of oxygen and $0.12 \mathrm{M}_{\odot}$ of iron are returned to the ISM for every $100 \mathrm{M}_{\odot}$ of star formation.
The net return of iron from Type Ia SNe, assuming the fiducial model parameters, is $0.17 \mathrm{M}_{\odot}$ for every $100 \mathrm{M}_{\odot}$ of star formation.
These nucleosynthetic yields are assumed to be independent of stellar metallicity.
In general yields will be metallicity dependent, however, for iron, the metallicity dependence is not expected to be strong \citep[e.g.][Figure 20]{andrews2017}.
A fixed fraction of stellar mass formed into stars is assumed to be recycled from the envelopes of CCSNe progenitors at its original metallicity and is referred to as the recycling parameter ($r=0.4$).
All free parameters excluding the star formation rate are assumed to be fixed with time\footnote{The fiducial model parameters are taken from a numerical implementation of the \citetalias{weinberg2017} model presented in \citet{andrews2017}, and are adopted here unless explicitly stated.
They are calibrated to reproduce the [O/Fe] versus [Fe/H] distribution of local thin disk, thick disk and halo stars.}.
Using these various approximations, it is possible to derive analytic equations for the evolution of chemical abundances. 
We refer interested readers to the original \citetalias{weinberg2017} paper for detailed derivations.

As discussed in \citetalias{weinberg2017}, the main governing parameters in their model are the mass-loading parameter (or outflow efficiency) $\eta$ and the star-formation efficiency, parameterized by the familiar gas-depletion timescale ($t_{\rm{dep}}=M_{\rm{gas}}/\dot{M}_{\ast}$).
The left-hand panel of Fig. \ref{fig:waf17_models} shows the predicted time evolution of \logz \ for different values of $\eta$ and $t_{\rm{dep}}$. 
Here, \logz \ refers to the iron abundance ([Fe/H]) in the \citetalias{weinberg2017} models.
It can clearly be seen that the assumed value of the mass-loading parameter has a pronounced effect, strongly influencing the typical \logz \ value reached after $\simeq 2.5$ \Gyr \ of evolution (corresponding to $z \simeq 2.5$). 
The reason is simply that by increasing the mass-loading parameter (from $\eta=3-10$ in our example) ISM enrichment is suppressed by the removal of more enriched gas from the galaxy.
In contrast, it is difficult to ascribe the low absolute abundances to a combination of relatively weak stellar winds ($\eta=3$) with long depletion timescales  (i.e. low star-formation efficiency, which will lead to lower metallicities simply due to the fact that fewer metals are being formed in stars).
As can be seen from Fig. \ref{fig:waf17_models}, even depletion timescales of $2$ \Gyr \ are not sufficient to reach agreement with the data. 
In reality, timescales of $\simeq 2$ \Gyr \ are more typical of local star-forming galaxies \citep[e.g.][]{leory2013}, and it has been found from observations of cold gas in high-redshift galaxies that the depletion timescale appears to decrease with increasing redshift, with typical values in the range $200 - 700$ \Myr \ across $z=1-6$ \citep[e.g.][]{tacconi2013,scoville2016,scoville2017}.
Therefore, within the context of the \citetalias{weinberg2017} model, the mass of metals produced assuming realistic values of the depletion timescale is too large to be consistent with the observations unless relatively large mass-loading parameters (i.e. high outflow efficiencies, $\eta \simeq 10$ on average across our sample) are assumed.
We note that the importance of strong outflows in shaping both the stellar and gas-phase metallicity is also found in many semi-analytic galaxy evolution models \citep[e.g][]{hirschmann2016,lian2018a}.

Further insights can be gained by investigating the mass dependence of the stellar metallicity. 
However, since the \citetalias{weinberg2017} model is only parameterized in terms of $\eta$, to simulate mass dependence a mapping between $\eta$ and mass is required.
Motivated by the outflow scaling relations from the FIRE simulations presented in \citet{muratov2015}\footnote{We note that these scaling relation have since been updated by \citet{hayward2017} and \citet{angles-alcazer2017} but we consider this simple parameterization sufficient for our purposes here.}, we assume $\eta$ is related to stellar mass via
\begin{equation}\label{eq:mass_loading_param_scaling}
\eta = \alpha\Big(\frac{M_{\ast}}{10^{10}\mathrm{M}_{\odot}}\Big)^{\beta},
\end{equation}
where $M_{\ast}$ is the galaxy stellar mass, $\alpha$ is the mass-loading parameter at $M_{\ast}=10^{10}\mathrm{M}_{\odot}$ and $\beta$ is the power-law exponent.
\citet{muratov2015} find a redshift-independent relation with $\alpha=3.55$ and $\beta=-0.351$.
Equation \ref{eq:mass_loading_param_scaling} can be easily incorporated into the \citetalias{weinberg2017} model in order evaluate the values of $\alpha$ and $\beta$ (and hence the $\eta(M_{\ast}$) relation) most consistent with our data.

The right-hand panel of Fig. \ref{fig:waf17_models} shows the difference in metallicity ($\Delta$\logz) as a function of stellar mass for different parameterizations of the stellar mass versus $\eta$ relationship assuming both the \citet{muratov2015} normalization ($\alpha=3.6$) and a slightly higher normalization implied by the \citetalias{weinberg2017} models (a value of $\alpha\simeq6$ is consistent with the highest mass - i.e. highest metallicity - data point in the left-hand panel of Fig. \ref{fig:waf17_models}).
The $\Delta$\logz \ values are calculated with respect to \logm \ $\simeq 9.56$.
The data favor a $\eta-M_{\ast}$ scaling relation with a similar power-law slope to the \citet{muratov2015} relation, irrespective of the absolute normalization.
A simple $\chi^2$ analysis returns a best-fitting power-law slope of $\beta=(-0.45, -0.40)$ for the $\alpha=(3.6, 6.0)$ cases respectively.
In either case the mass-loading parameter decreases by roughly a factor of 5 across the stellar mass range $10^8 < M_{\ast}/\mathrm{M}_{\odot} < 10^{10}$ with a median value of $\eta=8$ for $\alpha=3.6$ and $\eta=15$ for $\alpha=6.0$.
We note there is some degeneracy here between $\alpha$ and the assumed yields, however the corrections to the \citetalias{weinberg2017} would have to be quite large to make $\eta=3.0$ consistent with the observed data (e.g. \emph{both} the CCSNe and Type Ia SNe yields would have to be decreased by a factor 3).

In summary, whilst acknowledging the caveats associated with simple analytic models, we find that the \citetalias{weinberg2017} model supports a scenario in which the significantly sub-solar iron abundances of the galaxies in our sample, and the dependence of iron abundance on stellar mass, are a consequence of strong outflows which scale with \logm \ to the power $\simeq -0.4$.
In this scenario, the mass of gas outflowing from the galaxy, at $M_{\ast} = 10^9 \mathrm{M}_{\odot}$, is predicted to be roughly an order-of-magnitude greater than the gas mass being incorporated into new stars.
Finally we note that the inferences drawn from the \citetalias{weinberg2017} model are broadly similar to the those drawn from hydrodynamical simulations.
Although these methods are not fully independent, since simulations still rely some analytic sub-grid recipes, the agreement is encouraging and highlights the utility of simple analytic prescriptions for chemical evolution.

\subsection{Enhanced O/Fe ratios at high redshift}\label{subsec:o_fe_ratios}

    \begin{figure}
        \centerline{\includegraphics[width=\columnwidth]{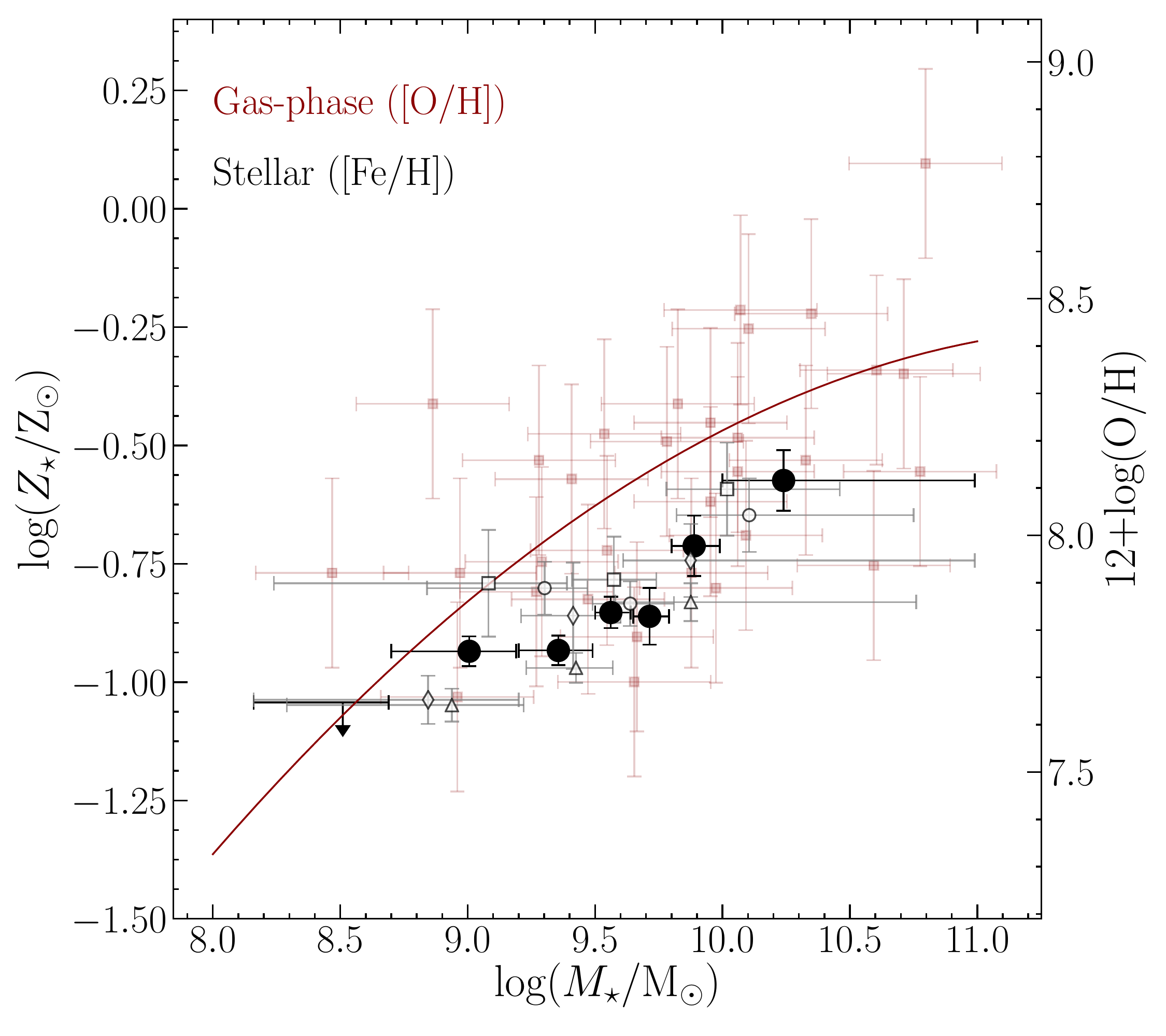}}
        \caption{A comparison of the stellar and gas-phase mass-metallicity relationships at $z\sim3.5$.
        The gas-phase data, tracing [O/H], are taken from the AMAZE/LSD survey of star-forming galaxies at $3<z<5$ with $\langle z \rangle = 3.4$ presented in \citet{troncoso2014}.
        The left-hand y-axis labels give metallicity in units of \logz \ while the right-hand labels give the equivalent 12+log(O/H) units commonly used for reporting gas-phase oxygen abundances.
        The red square data points with error bars show the individual galaxy data and the solid red curve is their best-fitting relationship.
        The black/grey data points are the stellar metallicities of our VANDELS sample plotted as in Fig. \ref{fig:normalized_mzr}.
        The comparison provides some evidence for alpha enhancement in star-forming galaxies at $z\sim3.5$ of the order (O/Fe) $\gtrsim 1.8$ $\times$ (O/Fe)$_{\odot}$.}
        \label{fig:alpha_enhancement}
    \end{figure}

Important insights can be gained by comparing the stellar metallicities of our sample to published gas-phase metallicities at similar redshifts.
The comparison is of interest because, while the stellar metallicities derived here are sensitive to elements dominating photospheric absorption in massive stars (namely iron), the gas-phase metallicities are determined from the nebular emission lines emitted from the surrounding \hii \ regions and are therefore sensitive to the dominant nebular coolants (namely oxygen).
Under the assumption that the chemical abundances of massive stars should be similar to the surrounding \hii \ region gas$-$out of which they presumably formed$-$then comparing stellar and gas-phase metallicities should be a good proxy for the O/Fe ratio in these galaxies \citep{steidel2016}.

Non-solar O/Fe ratios should be expected for galaxies with constant or rising star-formation histories and relatively young ages ($<500$ \Myr \ $-$ 1 \Gyr). 
This results from the fact that the release of Fe into the ISM is determined by both the rate of Type$-$Ia SNe and CCSNe, while the release of O (and other $\alpha$ elements) is determined only by the rate of CCSNe.
Crucially, the Type$-$Ia SNe rate is sensitive to the star-formation rate of a galaxy over a large range of epoch before the time of observations \citep[from $\simeq 50$ \Myr \ to $\simeq 10$ \Gyr \ see e.g.][]{maoz2012,dere_metallica}, while CCSNe enrich the ISM almost instantaneously (within a few \Myr \ of a given star-formation episode).
Therefore, for constant or rising SFHs, the abundance of the ISM will always be dominated by CCSNe products at young ages.
The yields of \citet{nomoto2006} predict, for a Salpeter IMF \citep{salpeter1955}, that CCSNe yields produce (O/Fe) $\simeq 4 - 6$ $\times$ (O/Fe)$_{\odot}$ for initial stellar metallicities $\simeq 0.1 - 2$ $\times$ $\mathrm{Z}_{\odot}$.
The yields of the \citet{chieffi2004} and \citet{limongi2006} fiducial model predict that CCSNe produce $1.5$M$_{\odot}$ of O and $0.12$M$_{\odot}$ of Fe for every $100$M$_{\odot}$ of star formation assuming a Kroupa IMF \citep{kroupa2001}, equivalent to (O/Fe) $\simeq 3$ $\times$ (O/Fe)$_{\odot}$.
Given the growing observational and theoretical evidence for constant or rising star-formation histories at the redshift of our sample \citep[e.g.][ see also Fig. \ref{fig:sim_sfr_chem}]{papovich2011,finlator2011,reddy2012b} we might therefore expect to see (O/H) $\simeq 3-6$ $\times$ (Fe/H).

Indeed, this result has already been reported for star-forming galaxies at $\langle z \rangle=2.3$ from the KBSS-MOSFIRE sample.
\citet{strom2018} present an analysis of the optical spectra of 148 galaxies allowing them to constrain [Fe/H] and [O/H] for individual galaxies, finding an average [O/Fe] $=0.42$ (i.e. (O/H) $\simeq 2.6$ $\times$ (Fe/H)), somewhat lower than the initial [O/Fe] $\simeq 0.7$ reported for a stack of 30 star-forming galaxies at the same redshift derived from a combined analysis of FUV and optical spectra presented in \citet{steidel2016}.
Nevertheless, both values are comparable to the expected range of [O/Fe] from the CCSNe yield models described above. 

Although we are unable to directly derive [O/H] for the galaxies in our sample, it is still instructive to compare the stellar metallicities reported here to [O/H] values at similar redshifts from the literature.
To this end we show, in Fig. \ref{fig:alpha_enhancement}, the gas-phase mass metallicity relationship for a sample of 34 galaxies at $3 < z < 5$ from the AMAZE/LSD survey \citep{maiolino2008,mannucci2009,troncoso2014} compared to the stellar metallicities derived for our VANDELS sample.
The AMAZE/LSD sample spans the stellar mass range $8.5 <$ \logm \ $< 11.0$ and is composed of Lyman-break selected galaxies with a bias to the most highly star-forming galaxies at a given mass \citep{maiolino2008}.
The metallicities were derived from a combination of the \oii, \hbeta \ and \oiii \ nebular emission lines using the semi-empirical metallicity calibration of \citet{maiolino2008} (M08).
Fig. \ref{fig:alpha_enhancement} shows that the stellar metallicities (Fe/H) of star-forming galaxies at $z\simeq3.5$ are systematically lower than the gas-phase metallicities (O/H) by $\simeq 0.25$ dex at all stellar masses compared to the mean relation found by \citet{troncoso2014}.
This value is comparable to the prediction of the \citetalias{weinberg2017} chemical evolution model described above, which, assuming the best-fitting $\eta-M_{\ast}$ relation, predicts [O/Fe] $\simeq0.2$ dex at $z=3.5$ (i.e. after $\simeq1.8$ \Gyr \ of evolution) across all stellar masses\footnote{This value is lower than that predicted directly from the yields because it incorporates the assumed star formation history and Type-Ia SNe DTD.}.

However, there are a number of caveats that should be acknowledged with respect to this comparison. 
Firstly, we note that the galaxies in the \citet{troncoso2014} sample have, on average, higher star-formation rates at fixed stellar mass than the galaxies in our sample (the median sSFR of the AMAZE/LSD galaxies is $5.4$ \Gyr$^{-1}$ compared to $4.4$ \Gyr$^{-1}$ for our sample).
Assuming a stellar mass-SFR-metallicity relation \citep[e.g.][]{mannucci2010} is in place at high redshifts (as claimed recently by \citet{sanders2018} at $z=2.3$) then, due to their higher star-formation rates, the AMAZE/LSD gas-phase metallicities will be biased low with respect to the true gas-phase metallicities of our sample.
In this case the offset of [O/Fe] $=0.25$ dex would be a lower limit.
Secondly, the comparison is affected by zero-point offsets in both the stellar and gas-phase metallicity values.
For example, in Appendix \ref{sec:bpassv2_comp} we show how using an alternative stellar population model \citep[in this case BPASSv2.2;][]{eldridge2017} can result in a systematic shift of $\simeq - 0.1$ dex in \logz.
Additionally, the zero-point offset in gas-phase metallicity calibrations, which has been well-studied in the local Universe \citep[e.g.][]{kewley2008}, implies that the calibration adopted by \citet{troncoso2014} (M08) returns metallicities intermediate between other calibrations with offsets of up to $\simeq \pm 0.2$ dex across the stellar mass range shown in Fig. \ref{fig:alpha_enhancement} \citep[e.g see Fig. 3 of][]{sanchez2019}.

Clearly, this comparison is currently limited to qualitative statements, and is hampered by the different selection biases affecting the VANDELS and AMAZE/LSD samples. 
A more definitive answer will have to wait until rest-frame optical observations of VANDELS galaxies are available, from which [O/H] can be directly estimated.
Nevertheless, adopting currently published values suggests that the O/Fe ratio is enhanced in high-redshift galaxies relative to the solar values, with (O/Fe) $\gtrsim 1.8$ $\times$ (O/Fe)$_{\odot}$.
One major practical implication of this result is that stellar population models which assume solar chemical abundance ratios are likely not valid for modeling the full spectra energy distribution of galaxies at high-redshift \citep[e.g.][]{steidel2016}\footnote{Although the Starburst99 models used here assume a solar abundance pattern, the fact that the FUV spectral features are primarily determined by Fe abundance means our recovered \logz \ values should not be strongly dependent on the detailed abundance pattern.
Nevertheless, in future it would clearly be desirable to test this assumption.}.

\section{Conclusions}\label{sec:conclusions}

In this paper we have presented the results of a study of the stellar mass-metallicity relationship for a sample of 681 star-forming galaxies at $2.3 < z < 5.0$ spanning the stellar masses range $8.2 <$ \logm \ $< 11.0$, based on deep rest-frame FUV spectroscopic data from the VANDELS survey \citep{mclure_vandels,pentericci_vandels}.
Stellar metallicities (to first order a proxy for the iron abundance) have been derived by fitting high-resolution theoretical Starburst99 WM-Basic stellar population synthesis models \citep{leitherer2010} to high signal-to-noise composite spectra in bins of stellar mass and redshift.
Our method is tested using both synthetic spectra generated from simulation data and FUV spectra of galaxies in the local Universe.
Finally, we have investigated the normalization and shape of the stellar mass-metallicity relationship at $2.3 < z < 5.0$, and compared our results to predictions from state-of-the-art cosmological simulations and simple one-zone analytic models for chemical evolution.
Our main results can be summarised as follows:

\begin{enumerate}

    \item We find a strong correlation between stellar metallicity and stellar mass at $2.3 < z < 5.0$ ($\langle z \rangle = 3.5$) with \logz \ monotonically increasing from $< -1.04$ for galaxies with $\langle M_{\ast} \rangle = 3.2 \times 10^{8} \rm{M}_{\odot}$ to $-0.57$ at $\langle M_{\ast} \rangle = 1.7 \times 10^{10} \rm{M}_{\odot}$.
    Across the full mass range  the metallicities (iron abundances) we derive are significantly sub-solar with $Z_{\ast} \lesssim 0.25 \mathrm{Z}_{\odot}$.

    \item We do not observe a strong relationship between stellar metallicity and redshift within our current sample.
    Within the error bars, the metallicities derived for composite spectra binned in redshift and stellar mass are consistent with scattering around the stellar-mass only relationship.
    
    \item The stellar mass-metallicity relation we derive at $\langle z \rangle = 3.5$ is offset to lower metallicities by $\simeq 0.6$ dex compated to the $z=0$ relation for star-forming galaxies derived from optical spectroscopic data \citep{zahid2017}.
    There are some systematic uncertainties related to this comparison (e.g. probing different stellar populations and/or abundance types), but similar results are obtained by comparing to the FUV-based metallicities at $z=0$ from FOS-GHRS data. 

    \item The normalization of the $\langle z \rangle = 3.5$ stellar mass-metallicity relation is not well reproduced by simulations.
    Absolute metallicity values tend to fall either systematically above (e.g \fiby) or below (e.g. \simba, FIRE) our data.
    However, given the relatively large uncertainties in stellar yields, it is hard to draw strong conclusions from this observation.
    On the other hand, the shape of the relation is generally well recovered.
    We postulate that this is likely due to the fact that the scaling between outflow efficiency (parameterized by the mass loading parameter $\eta = \dot{M}_{\rm{outflow}}/\dot{M}_{\ast}$) and stellar mass is similar across all of the simulations.

    \item We investigate the parameters affecting the shape and normalization of the stellar mass-metallicity relation using detailed one-zone analytic chemical evolution models \citep[the \citetalias{weinberg2017} models;][]{andrews2017,weinberg2017}. 
    We find that these models can reproduce the observed shape the relation by assuming a power-law scaling relation between $\eta$ and $M_{\ast}$  that is similar to that derived from fully cosmological simulations \citep[e.g.][]{muratov2015} and predicted from analytic models of momentum-driven winds \citep{hayward2017}.
    A comparison of the models and simulations to our data suggest that $\eta \propto M_{\ast}^{\beta}$ with $\beta \simeq -0.4$.
    
    \item Furthermore, simulations which include state-of-the-art prescriptions for stellar feedback (FIRE and \simba), as well as the \citetalias{weinberg2017} model, suggest that an average mass-loading parameter of $\langle \eta \rangle \simeq 10$ is required for consistency with our data, ranging from $\eta \simeq 6$ at $M_{\ast} = 10^{10} \rm{M}_{\odot}$, to $\eta \simeq 40$ at $M_{\ast} = 10^{8} \rm{M}_{\odot}$.
    Although the absolute normalization of the $\eta-M_{\ast}$ relation is subject to yield uncertainties, the reasonable agreement with predictions from cosmological simulations is encouraging.

    \item Finally, by comparing the stellar metallicities derived here (which trace Fe/H)  to published gas-phase metallicities at similar redshifts \citep{troncoso2014}, we find that the gas-phase abundances (which trace O/H) are likely enhanced by at least a factor of $\simeq 2$.
    This comparison provides further evidence that star-forming galaxies at these epochs are $\alpha$-enhanced systems, although it will require direct [O/H] estimates for the galaxies in our sample to provide a more robust measurement.

\end{enumerate}

\section{Acknowledgments}

FC, RJM, JSD, SK, AC and DJM acknowledge the support of the UK Science and Technology Facilities Council.
This work is based on data products from observations made with ESO Telescopes at La Silla Paranal Observatory under ESO programme ID 194.A-2003(E-Q).
AC acknowledges the grants PRIN MIUR 2015, ASI n.I/023/12/0 and ASI n.2018-23-HH.0
GC has been supported by the INAF PRIN-SKA 2017 program 1.05.01.88.04.
This research made use of Astropy, a community-developed core Python package for Astronomy \citep{astropy2013}, NumPy and SciPy \citep{oliphant2007}, Matplotlib \citep{hunter2007}, {IPython} \citep{perez2007} and NASA's Astrophysics Data System Bibliographic Services.

\bibliographystyle{mnras}                      
\bibliography{vandels_stellar_metallicities}       

\appendix

\section{FOS-GHRS local sample}

\begin{table}
    \centering
    \caption{Sample of local starbursts and star-forming galaxies taken from \citet{leitherer2011}. We use the M/L relations from \citet{mcgaugh2014} to convert from $M_{\rm{K}}$ to \logm.}\label{table:fos_ghrs_sample_appdx}
    \begin{tabular}{ccc}
        \hline
        \hline
        Galaxy & $M_{\rm{K}}$ & \logm\\
        \hline
        NGC2363-a & $-$17.1 & 7.93\\
        NGC2363-b & $-$17.1 & 7.93\\
        NGC1705 & $-$18.0 & 8.29\\
        SBS0335-052 & $-$18.0 & 8.29\\
        NGC1569 & $-$18.9 & 8.65\\
        NGC4214 & $-$19.4 & 8.85\\
        NCG5253-UV1 & $-$19.5 & 8.89\\
        NCG5253-UV2 & $-$19.5 & 8.89\\
        \hline
        He2-10a & $-$21.1 & 9.53\\
        He2-10b & $-$21.1 & 9.53\\
        NGC4670 & $-$21.4 & 9.65\\
        NGC1741 & $-$21.8 & 9.81\\
        IC3639 & $-$23.0 & 10.29\\
        NGC7714 & $-$23.2 & 10.37\\
        NGC5457-Searle5 & $-$23.8 & 10.61\\
        NGC7552 & $-$24.2 & 10.77\\
        NGC4038-405 & $-$24.5 & 10.89\\
        NGC4038-442 & $-$24.5 & 10.89\\
        \hline
        NGC3690 & $-$25.0 & 11.09\\
        NGC5135 & $-$25.0 & 11.09\\
        NGC7130 & $-$25.0 & 11.09\\
        NGC1068-pos1 & $-$25.1 & 11.13\\
        NGC1068-pos3 & $-$25.1 & 11.13\\
        NGC1068-pos5 & $-$25.1 & 11.13\\
        NGC1068-pos8a & $-$25.1 & 11.13\\
        \hline
    \end{tabular}
\end{table} 

    \begin{figure*}
        \centerline{\includegraphics[width=6.0in]{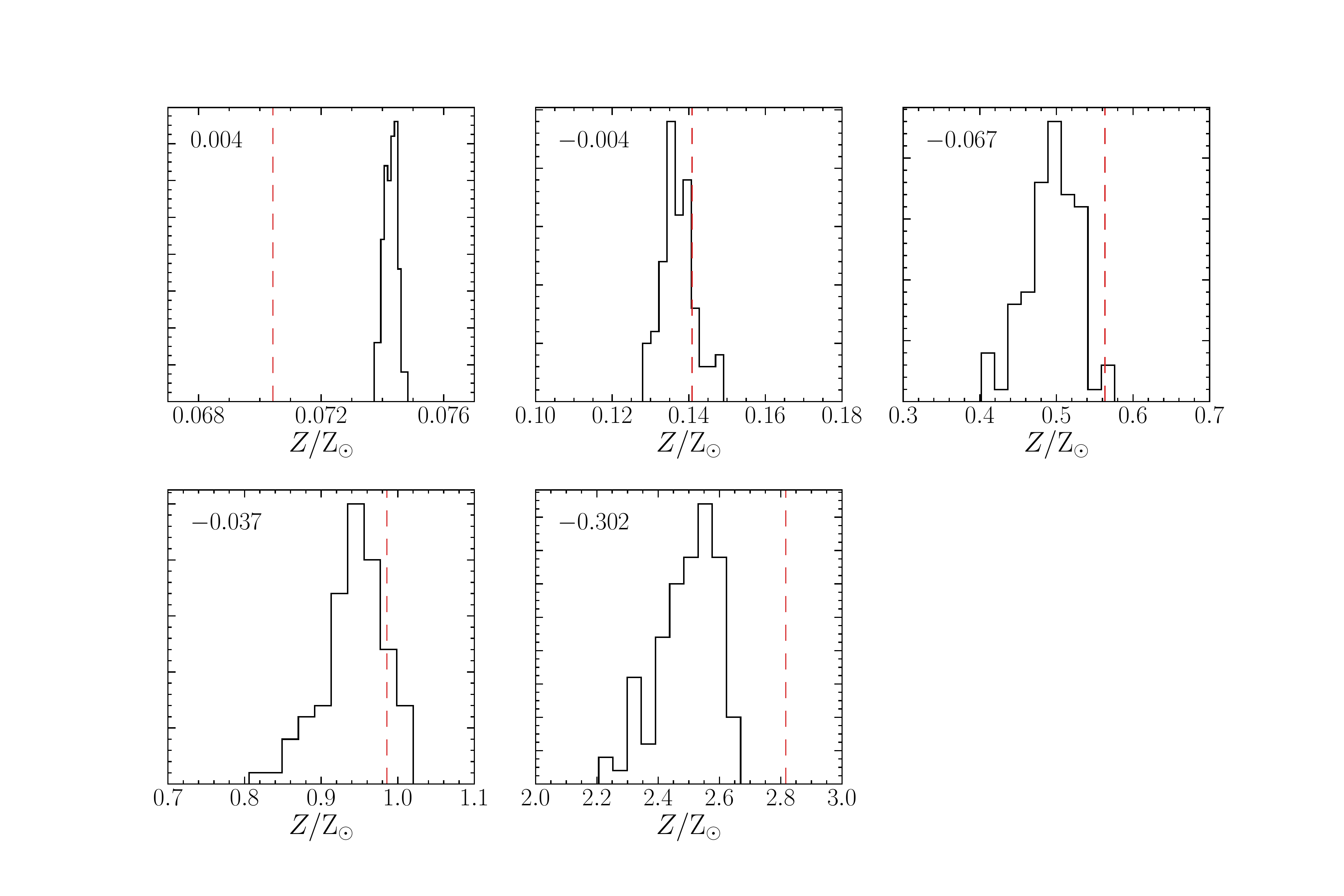}}
        \caption{The distribution of recovered metallicity for each of the five default Starburst99 WM-Basic template spectra ($Z_{\ast}=0.001, 0.002, 0.008, 0.014, 0.040$) after mimicking the effect of redshift uncertainties when creating composite spectra (see text for details).
        Each panel shows one of the five default templates with the true metallicity indicated by the dashed vertical red line.
        The distribution of recovered metallicities is shown in black.
        The offset between the median recovered metallicity and the true metallicity is quoted in the top left-hand corner of each panel.}
        \label{fig:redshift_uncertainties}
    \end{figure*}

The local FUV spectra are taken from a complication of 28 local starbursts and star-forming galaxies presented in \citet{leitherer2011} \citepalias{leitherer2011}.
\citetalias{leitherer2011} present 46 rest-frame UV spectra observed with the Faint Object Spectrograph (FOS) and the Goddard High Resolution Spectrograph (GHRS) of the \emph{HST}.
The spectral resolution of the individual spectra spans the range $0.5-3\rm{\AA}$ depending on the instrument, aperture size and physical extent of the object being observed.
For consistency with the the VANDELS data, we smooth all spectra to a common $3\rm{\AA}$ resolution.
We also require that the spectra have wavelength coverage in the interval $1410 \leq \lambda \leq 1450 \rm{\AA}$ to enable an analysis of normalized composite spectra, and finally that the corresponding galaxy has a measurement of absolute K-band magnitude which we can use to estimate the stellar mass.
This selection leaves 26/46 of the original spectra from 18/28 of the original local starbursts and star-forming galaxies presented in \citetalias{leitherer2011}.
A list of the individual spectra used in this work is presented in Table \ref{table:fos_ghrs_sample_appdx}. 

\section{Effect of redshift uncertainties}\label{sec:app_redshift_uncertainties}

One potential source of bias in our analysis relates to the effect of redshift uncertainties when stacking.
The redshifts for the individual VANDELS spectra are constrained primarily by the strong interstellar absorption features and/or \lya \ emission line.
These features originate from outflowing gas and are therefore often redshifted or blueshifted by up to hundreds of kms$^{-1}$ with respect to the systemic redshift of the galaxy.
For example, from a sample of $89$ galaxies at $z\sim2-3$, \citet{steidel2010} find an average velocity offset for redshifts derived from interstellar absorption lines of $\langle \Delta v_{\mathrm{IS}} \rangle = -164 \pm 16$ kms$^{-1}$, and of $\langle \Delta v_{\mathrm{Ly\alpha}} \rangle = +445 \pm 27$ kms$^{-1}$ for redshifts derived from \lya \ emission.
Each galaxy in our sample will therefore have some unknown redshift offset which could potentially result in a systematic bias when deriving metallicities.
In particular, the weak continuum features which are used to constrain the metallicity via continuum fitting could be `washed out' to make the continuum look increasingly featureless, and therefore bias measurements towards lower metallicity solutions.

We tested the effect of redshift uncertainties using simulated galaxy spectra.
For each of the five default Starburst99 WM-Basic templates ($Z_{\ast}=0.001, 0.002, 0.008, 0.014, 0.040$) we produced 100 versions (i.e. roughly the number of galaxies in our mass stacks) each with a random velocity offset distributed uniformly within the range $-150 < \Delta v  < 500$ kms$^{-1}$.
The 100 spectra were then median combined to mimic stacking spectra with random, unknown, redshift uncertainties.
These spectra were convolved to the resolution of the VANDELS data and gaussian noise was added assuming a S/N per pixel of 15. 
The metallicity was then derived using the method applied to the observed data as described in Section \ref{sec:method}.
For each template the process was repeated 100 times and the distribution of recovered metallicities is shown in Fig. \ref{fig:redshift_uncertainties}.
It can be seen that although there is a systematic offset to lower metallicities for all but the lowest metallicity template (for which it is impossible to measure a lower metallicity), the effect is relatively small ($\leq 10 \%$).
We therefore conclude that the low metallicities derived for our VANDELS sample are robust against the effect of redshift uncertainties.

\section{BPASSv2.1 stellar population synthesis models}\label{sec:bpassv2_comp}

In our main analysis we focused on using the Starburst99 WM-Basic stellar population synthesis models to fit the VANDELS spectra for the reasons outlined in Section \ref{sec:method}.
However, we have also performed the same analysis using the BPASSv2.1
 stellar population synthesis models described in \citet{eldridge2017}.
It is well known that using different models will cause systematic offsets in derived parameters \citep[e.g.][]{conroy_review,cidfernandes2014} and it is useful to have an estimate of how significant this offset my might be in the case of FUV-derived metallicities. 

The unique feature of the BPASSv2.1 models is the inclusion of massive binary star evolution, which can have a strong effect on the predicted UV spectrum, particularly at low metallicities.
For example, at $Z_{\ast}/\mathrm{Z}_{\odot} < 0.35$, a phenomenon known as `quasi-homogeneous evolution' can occur \citep[QHE; e.g.][]{eldridge2012}, which results in stars living longer on the main-sequence and becoming hotter than single stars at similar mass and metallicity.
QHE can have a profound effect on the estimated ages and ionizing photon output of stellar populations.
Stellar metallicities should be less affected, particularly since the O-star models in BPASSv2.1 are also generated with WM-Basic code, albeit with a different parameter set to Starburst99.
Nevertheless, it worth testing this assumption.

	\begin{figure}
        \centerline{\includegraphics[width=\columnwidth]{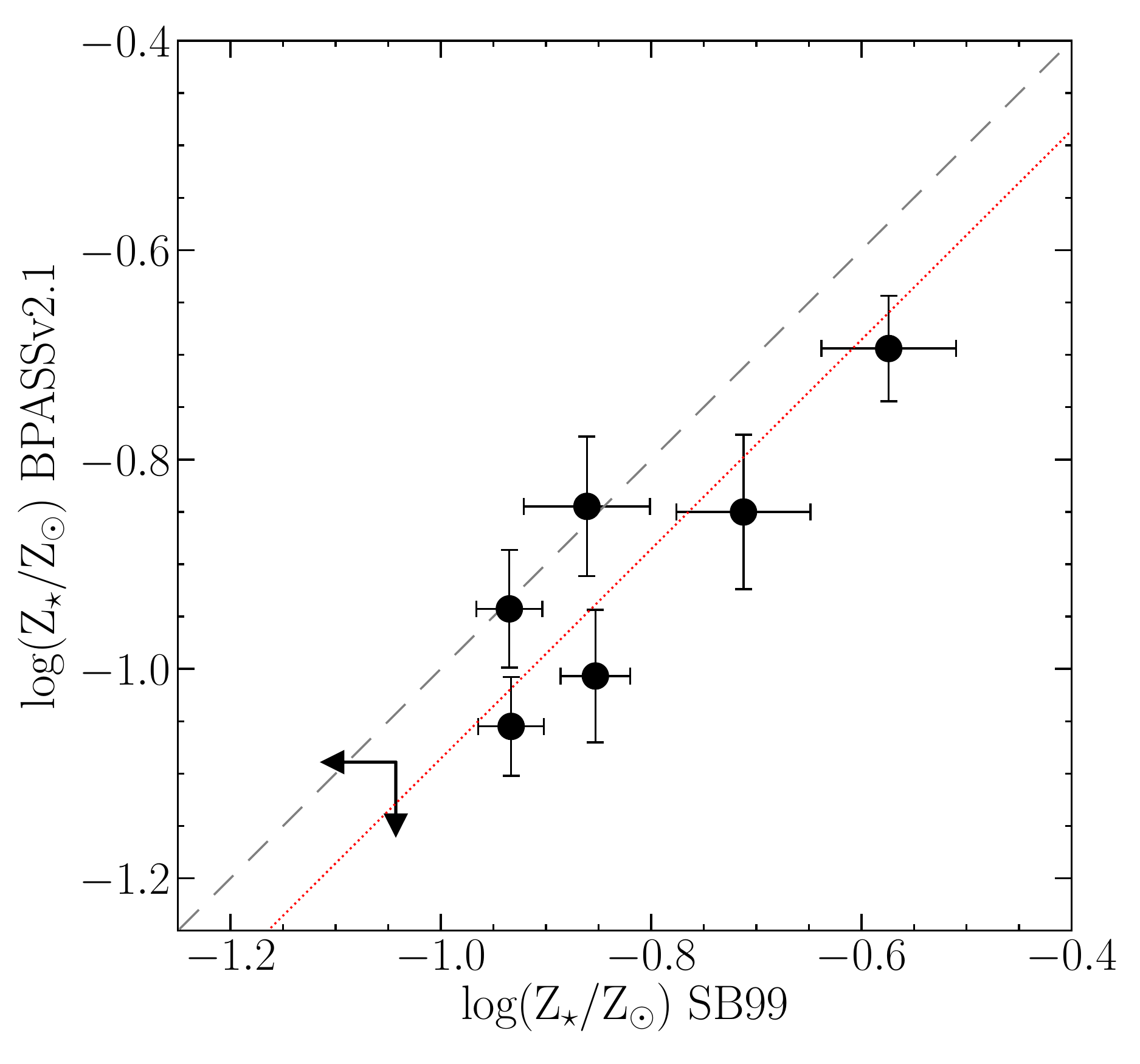}}
        \caption{A comparison between stellar metallicities derived from the Starburst99 WM-Basic and BPASSv2.1 models for the VANDELS mass-stacks.
        The grey dashed line shows the 1:1 relation and the red dotted line shows the best-fitting constant offset from the 1:1 line.
        Both estimates are clearly correlated with the BPASSv2.1 metallicities offset to slightly lower metallicity values by $\sim 0.09$ dex.
        For both Starburst99 and BPASSv2.1, the metallicity derived for the lowest mass bin is an upper limit (indicated by arrows in the figure).
        The figure illustrates that adopting the BPASSv2.1 metallicity values as opposed to the Starburst99 values would not affect our main results.}
        \label{fig:bpass_comp}
    \end{figure}

We considered a 100 \Myr \ constant star formation BPASSv2.1 model which included binary star evolution and had an IMF cutoff of 100 M$_{\odot}$, with IMF index of $-1.3$ between $0.1 - 0.5$ M$_{\odot}$ and $-2.35$ above 0.5 M$_{\odot}$.
The model is referred to as BPASSv2.1-100bin.
We considered the following set of metallicity values $Z_{\ast} =$
(0.001, 0.002, 0.003, 0.004, 0.006, 0.008, 0.010, 0.014, 0.020,
0.030, 0.040), interpolating between the models in the same way as for Starburst99.
Each of the seven mass stacks were fitted with the BPASSv2.1-100bin model in the same way as was done{} for the Starburst99 models (see Section \ref{sec:method} for details).
Fig. \ref{fig:bpass_comp} compares the metallicities derived from the BPASSv2.1-100bin model and the metallicities derived from Starburst99.
It can be seen that the metallicity estimates are clearly correlated, with the BPASSv2.1 values offset to slightly lower metallicity by $\sim 0.09$ dex (roughly a factor of 1.2).
However, it is clear from this figure that adopting the BPASSv2.1 model estimates would not affect our main results.

\bigskip

\noindent
$^{1}$SUPA\thanks{Scottish Universities Physics Alliance}, Institute for Astronomy, University of Edinburgh, Royal Observatory, Edinburgh EH9 3HJ\\
$^{2}$University of the Western Cape, Bellville, Cape Town 7535, South Africa\\
$^{3}$South African Astronomical Observatories, Observatory, Cape Town 7925, South Africa\\
$^{4}$Instituto de Investigaci\'on Multidisciplinar en Ciencia y Tecnolog\'ia, Universidad de La Serena, Ra\'ul Bitr\'an 1305, La Serena, Chile\\
$^{5}$Departamento de F\'isica y Astronom\'ia, Universidad de La Serena, Av. Juan Cisternas 1200 Norte, La Serena, Chile\\
$^{6}$INAF - Osservatorio Astronomico di Bologna, via P. Gobetti 93/3,
I-40129, Bologna, Italy\\
$^{7}$INAF$-$Osservatorio Astronomico di Roma, Via Frascati 33, I-00040 Monte Porzio Catone (RM), Italy\\
$^{8}$University of Bologna, Department of Physics and Astronomy (DIFA) 
Via Gobetti 93/2- 40129, Bologna, Italy\\
$^{9}$INAF - Osservatorio Astrofisico di Arcetri, Largo E. Fermi 5, I-50125, Firenze, Italy\\
$^{10}$European Southern Observatory, Karl-Schwarzschild-Str. 2, 86748 Garching b. M\"unchen, Germany\\
$^{11}$The Cosmic Dawn Center, Niels Bohr Institute, University of Copenhagen, Juliane Maries Vej 30, DK-2100 Copenhagen {{\O}}, Denmark\\
$^{12}$INAF-Astronomical Observatory of Trieste, via G.B. Tiepolo 11, 34143 Trieste, Italy\\
$^{13}$INAF-IASF Milano, via Bassini 15, I-20133, Milano, Italy\\
$^{14}$ N\'ucleo de Astronom\'ia, Facultad de Ingenier\'ia, Universidad Diego Portales, Av. Ej\'ercito 441, Santiago, Chile\\
$^{15}$Space Telescope Science Institute, 3700 San Martin Drive, Baltimore, MD 21218, USA\\
$^{16}$European Southern Observatory (Chile)\\
$^{17}$Department of Physics and Astronomy, University of California, Los Angeles, 430 Portola Plaza, Los Angeles, CA 90095, USA

\label{lastpage}
\bsp
\end{document}